\documentclass[prd,nofootinbib,twocolumn,superscriptaddress,preprintnumbers,balancelastpage]{revtex4-1}

\pdfoutput=1
\usepackage{amsmath,amssymb}
\usepackage{amsfonts}
\usepackage{bm,bbm}
\usepackage{graphicx}
\usepackage{mathrsfs}
\usepackage{slashed}
\usepackage{booktabs}
\usepackage{tabu}
\usepackage[dvipsnames]{xcolor}
\usepackage[normalem]{ulem}
\usepackage{tikz}
\usepackage{soul}

\usepackage{hyperref}
\hypersetup{colorlinks,citecolor= blue,linkcolor= blue, urlcolor=blue}

\usepackage{xcolor}
\usepackage{ulem}
\usepackage{array}
\usepackage{verbatim}
\usepackage{epsfig}
\usepackage{multirow}

\newcommand{\be}{\begin{equation}}
\newcommand{\ee}{\end{equation}}
\newcommand{\ba}{\begin{array}}
\newcommand{\ea}{\end{array}}
\newcommand{\bea}{\begin{eqnarray}}
\newcommand{\eea}{\end{eqnarray}}

\usepackage{ulem,fancyvrb}
\usepackage{xcolor}

\definecolor{blue-violet}{rgb}{0.54, 0.17, 0.89}
\definecolor{amethyst}{rgb}{0.6, 0.4, 0.8}

\begin{document}

\title{Search for strongly interacting dark matter at Belle II}

\author{Jinhan Liang}\email{jinhanliang@m.scnu.edu.cn}
\affiliation{Key Laboratory of Atomic and Subatomic Structure and Quantum Control (MOE), Guangdong Basic Research Center of Excellence for Structure and Fundamental Interactions of Matter, Institute of Quantum Matter, South China Normal University, Guangzhou 510006, China}
\affiliation{Guangdong-Hong Kong Joint Laboratory of Quantum Matter, Guangdong Provincial Key Laboratory of Nuclear Science, Southern Nuclear Science Computing Center, South China Normal University, Guangzhou 510006, China }

\author{Zuowei Liu}\email{zuoweiliu@nju.edu.cn}
\affiliation{Department of Physics, Nanjing University, Nanjing 210093, China}

\author{Lan Yang}\email{lanyang@smail.nju.edu.cn}
\affiliation{Department of Physics, Nanjing University, Nanjing 210093, China}

\begin{abstract}

A small component of dark matter (DM) that is strongly interacting with the standard model sector is consistent with various experimental observations. Despite the small abundance, strongly-interacting DM can lead to pronounced signals in DM direct detection experiments. We study Belle II sensitivity on strongly-interacting DM that has a MeV-GeV mass and couples with electrons. By taking into account the substantial interactions between DM and electrons within detectors, we compute the ``ceiling'' of the mono-photon signature at Belle II, beyond which the mono-photon channel loses its sensitivity, and visible ECL clusters due to DM scatterings assume significance. We study two ECL signatures for strongly-interacting DM: the mono-cluster and the di-cluster channels. To carry out detailed calculations and to compare with other constraints, we consider DM models with light mediators, as they naturally lead to sizable interaction cross sections. We compute exclusion regions for the di-cluster, mono-cluster, and mono-photon channels. We find that Belle II (with currently accumulated data of 362 fb$^{-1}$) can rule out a significant portion of the parameter space above the ceilings of the constraints from various DM direct detection and neutrino experiments, for the vector mediator case with mass $\gtrsim 10$ MeV. Belle II also offers superior constraints on new light particles compared to PBH for the scalar mediator with mass $\gtrsim 10$ MeV.

\end{abstract}

\maketitle

\section{Introduction}

Dark matter (DM) is usually assumed to have a weak interaction 
cross section with standard model (SM) particles 
\cite{Bertone:2016nfn, Arbey:2021gdg, Bertone:2004pz}. 
Nonetheless, strongly-interacting DM  
that has a significant interaction cross section with 
the SM sector 
is allowed if it only constitutes a small DM abundance   
\cite{Starkman:1990nj, McGuire:2001qj, Mack:2007xj}. 
One of the intriguing aspects of strongly-interacting DM lies in its potential to significantly increase the velocity of DM in astrophysical environments, thereby enhancing signals in DM direct detection (DMDD) experiments.
Notable scenarios include up-scatterings induced by 
cosmic rays  
\cite{Cappiello:2018hsu, Bringmann:2018cvk, Ema:2018bih}, 
diffuse supernova neutrinos \cite{Das:2021lcr}, 
and blazars \cite{Wang:2021jic}.

A small component of DM can 
be strongly interacting with the SM sector
because various experimental constraints 
can be alleviated due to either the strong interaction 
cross section or the small DM abundance. 
For example, the flux of strongly-interacting DM at 
underground DMDD experiments is  
shielded by the overburden. 
The constraints from DM indirect detection 
are limited by the small DM abundance. 
The CMB constraints start to lose sensitivity when 
the DM fraction becomes less than 0.4\% \cite{Boddy:2018wzy}.

Particle colliders present a great opportunity for strongly-interacting DM 
searches, as they are not impeded by the strong absorption effects 
and the small DM abundance. 
The conventional DM signature at colliders is 
the missing momentum search
\cite{Birkedal:2004xn, Feng:2005gj, Beltran:2010ww, Bai:2010hh}, 
which, however, is not applicable to strongly-interacting DM. 
This is because strongly-interacting DM has 
a substantial interaction cross section with the 
detectors, 
and is therefore more likely to scatter with the detectors, 
leading to visible collider signatures, 
such as trackless jet signals at the LHC 
\cite{Bai:2011wy, Daci:2015hca, Bauer:2020nld}. 
Thus, in the missing momentum search,
there exists a ``ceiling'' on the DM-SM interaction cross section, 
beyond which DM is no longer considered  invisible at colliders 
\cite{Daci:2015hca}.

As previous investigations have primarily focused on hadron collider 
signals arising from strongly-interacting DM, 
we study its signatures at electron-positron colliders in this paper. 
This is particularly relevant for sub-GeV DM that 
interacts with electrons. 
Sub-GeV DM is less likely to deposit 
significant nuclear recoil energy in many of the leading 
underground DMDD experiments that aim to detect 
weakly interacting massive particles (WIMPs). 
Consequently, electron recoils assume significance
in DMDD experiments for sub-GeV DM, 
because of the small electron mass 
\cite{Essig:2011nj,Essig:2012yx}. 
The study of electron collider constraints on 
sub-GeV electro-philic DM becomes imperative for a comprehensive understanding.

Thus, we study constraints on strongly-interacting DM 
from the Belle II experiment, which is operated at 
$\sqrt{s}=10.58$ GeV and is an ideal experiment  
to detect DM within the MeV-GeV mass range that interacts with electrons 
\cite{Hauth:2018fgp,Liang:2019zkb,Mohlabeng:2019vrz, Duerr:2019dmv,Filimonova:2019tuy,
Kang:2021oes, Liang:2021kgw, Liang:2022pul, Duerr:2020muu,Acanfora:2023gzr}. 
As DM is neutral and has no track in the central drift chamber (CDC), 
our detection strategy relies on the electromagnetic calorimeter (ECL) detector at Belle II. 
With a substantial interaction with the ECL, 
DM can scatter with atomic electrons in the ECL, 
resulting in recoiled electrons with significant energy. 
Subsequently, these recoiled electrons generate electromagnetic (EM) showers, 
giving rise to distinct ``cluster'' signatures in the ECL. 
We analyze two types of DM-induced cluster signals at Belle II: 
the mono-cluster and the di-cluster. 
We then compute exclusion regions 
in the parameter space spanned by 
the DM-electron interaction cross section 
and the DM mass, 
for three DM signatures, 
including the mono-cluster, the di-cluster, 
and the mono-photon channels. 
We find that the di-cluster channel typically probes the parameter 
space 
above the ceiling of the mono-photon channel. 
In contrast, the mono-cluster channel rarely extends into 
new parameter space beyond that probed by the mono-photon channel.

To compare Belle II sensitivity with various 
constraints from 
DMDD and neutrino experiments, 
we consider DM models with 
light mediators that couple to both DM and electrons. 
We find that Belle II can rule out 
a significant portion of the parameter space 
above the ceilings of the constraints from various DMDD 
\cite{Essig:2017kqs, Aprile:2019xxb, An:2017ojc, Emken:2019tni} 
and neutrino experiments 
\cite{Calabrese:2022rfa, Dent:2020syp}, 
for the vector mediator case with mass $\gtrsim 10$ MeV.
For the ultralight mediator case, however, 
the parameter space probed by Belle II 
and that probed by DMDD experiments 
are well separated. 
We also find that Belle II constraints on 
beyond-the-SM (BSM) particles 
are better than those from primordial black holes (PBH), 
for the scalar mediator 
with mass $\gtrsim 10$ MeV.

The rest of the paper is organized as follows. 
In section \ref{sec:int}, we provide a brief discussion of
the DM-induced signatures at Belle II. 
We then discuss in detail three different DM-induced signatures 
at Belle II: 
the di-cluster channel in section \ref{sec:dicluster}, 
the mono-photon channel in section \ref{sec:mono-photon}, 
and the mono-cluster channel in section \ref{sec:monocluster}. 
To compute Belle II sensitivity on strongly-interacting DM, 
we consider DM models with four different types of 
light mediators in section \ref{sec:model}. 
We compute experimental constraints on the light mediators 
in section \ref{sec:otherexp}, 
including electron $g-2$, 
electron beam dump, BaBar, 
and {M\o ller} scattering. 
We compute Belle II sensitivities 
on the DM models with light mediator models 
in section \ref{sec:CPBelle2},  
and further compare them 
with constraints from DMDD and neutrino experiments in section \ref{sec:refXsec}. 
We also compare Belle II sensitivities on BSM particles 
with PBH constraints in section \ref{sec:PBH}. 
We summarize our findings in section \ref{sec:summary}. 
In appendix \ref{sec:CSequation}, we provide the cross sections 
of the 
$e^+e^-\to\chi\bar{\chi}$, 
$e^+e^-\to\chi\bar{\chi}\gamma$, and 
$\chi e^-\to\chi e^-$ 
processes, 
for the four different mediator models.

\section{ECL response of DM interaction}
\label{sec:int}

In this section we discuss the detector response 
to DM interactions with electrons. 
We focus on the ECL detector at Belle II, 
which is composed of CsI crystals. 
The electrons in the ECL can receive  
significant recoil energy when hit by 
the incident DM. 
The recoiled electron can subsequently 
generate EM 
showers in the ECL,  
leading to the so-called ``cluster'' signature if the 
total energy deposited across multiple cells 
exceeds 20 MeV \cite{Shinya2011}. 
We note that this energy threshold is similar to 
the critical energy of the ECL, 
which is $E_c = 20.7$ MeV \cite{Grupen:2008zz}.

To intuitively estimate the interaction rate with the ECL, 
we define the mean free path of DM 
in ECL as follows 
\be
\lambda_c =\frac{1}{n_E \sigma_{c}},
\ee
where $n_E$ is the electron number density of the ECL, 
and $\sigma_c$ is the DM-electron cross section 
with the electron recoil energy $E_r>E_c$.
Thus, it is expected that one 
has to seriously take into account 
interactions between DM and electrons within detectors 
for DM searches at Belle II, 
when $\lambda_c \lesssim L_E$,
where $L_E$ is the length of the ECL detector. 
In the following, 
we investigate how these DM-detector interactions  
invalidate the missing momentum signature and 
give rise to new visible DM signatures within the ECL. 
By comparing with actual calculations, 
we find that the $\lambda_c \lesssim L_E$ condition
serves as a fairly good criterion for determining the 
ceilings of the missing momentum signature.

\section{di-cluster signal at Belle II}
\label{sec:dicluster}

\begin{figure}[htbp]
\includegraphics[width=0.35 \textwidth]{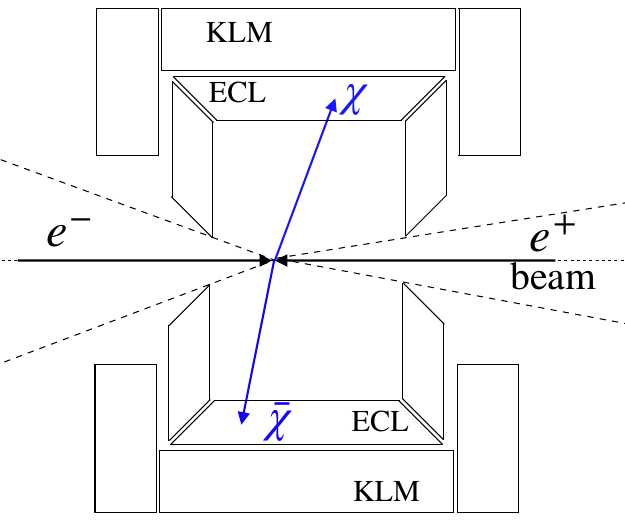}
\caption{Schematic view of the signal event 
in the di-cluster channel at the Belle II detector.}
\label{fig:di-cluster}
\end{figure}

DM can be pair-produced at Belle II 
via the $e^+ e^- \to \chi \bar{\chi}$ process with $\chi$ denoting the DM particle, 
as illustrated in Fig.~(\ref{fig:di-cluster}).
Because DM is neutral,
it leaves no track in the CDC.
If the interaction between DM and the ECL 
is sufficiently strong, 
both $\chi$ and $\bar{\chi}$ can result in clusters in the ECL. 
Thus, the observable signal for SIDM consists 
of two back-to-back trackless clusters 
in the center of mass (c.m.) frame; 
we denote this as the di-cluster signature. 
Note that the experimental reconstruction of 
two back-to-back trackless clusters in the c.m.\ frame may encounter challenges. 
This is attributed to the potential scenario where only partial energy 
of DM is deposited in the ECL. 
In this case, the two trackless clusters 
are back-to-back in the transverse plane.

\begin{figure}[htbp]
\includegraphics[width=0.45 \textwidth]{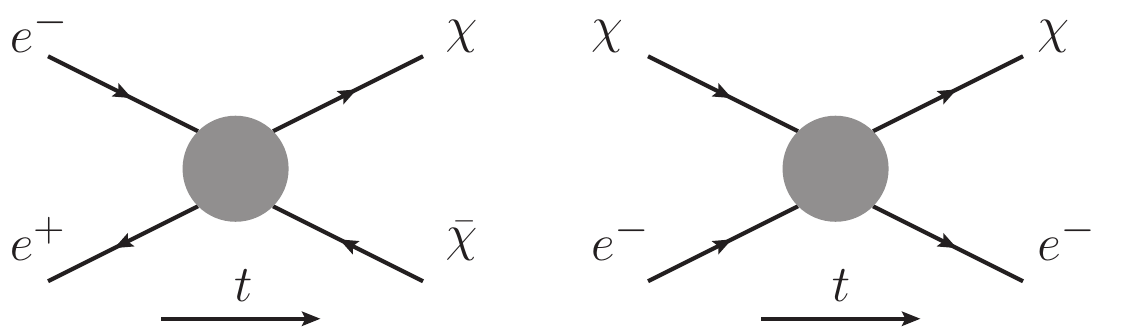}
\caption{Diagrams for DM production (left) and DM-electron scattering (right).}
\label{fig:DM:diagram}
\end{figure}

In our analysis, 
we only consider the final state DM particles in  
the barrel region of the ECL detector, which has  
better hermiticity compared to endcap regions \cite{Belle:2000cnh}. 
The barrel region of the ECL detector 
has a polar angle of $32.2^\circ<\theta<128.7^\circ$ 
in the lab frame. 
We compute the DM-induced 
di-cluster signal events in the barrel region of the ECL via
\begin{equation}
N_s = \mathcal{L}\int dz_{\chi}^* 
\frac{d\sigma_{\chi\bar{\chi}}}{dz_{\chi}^*} 
P_\chi (z_{\chi}^*)
P_{\bar{\chi}}(-z_{\chi}^*), 
\label{eq:di-chi}
\end{equation}
where 
$\mathcal{L}$ is the integrated luminosity at Belle II, 
$z_{\chi}^*=\cos \theta_{\chi}^*$
with $\theta_{\chi}^*$ being the polar angle of $\chi$ in 
the c.m.\ frame, 
$d \sigma_{\chi\bar{\chi}}/dz_\chi^*$ is 
the differential cross section for 
the $e^+ e^- \to \chi \bar{\chi}$ process,
and $P_{\bar\chi}$ ($P_{\chi}$) is the interaction probability 
between DM $\bar\chi$ (${\chi}$) and the ECL. 
Here, we have used the fact that 
$z_{\bar \chi}^* \equiv \cos \theta_{\bar \chi}^* 
= -z_\chi^*$ 
in the $e^+ e^- \to \bar \chi \chi$ process. 
Hereafter, for a kinematic variable $A$ in the lab frame, 
we use $A^*$ to denote its value in the c.m.\ frame.

Because Belle II collides 
a 7 GeV electron with a 4 GeV positron \cite{Belle-II:2018jsg}, 
the polar angle of $\chi$ (or $\bar \chi$) 
in the lab frame is related to 
its value in the c.m.\  frame via 
\begin{equation}
  z = f(z^*) = \frac{ z^*  \beta_\chi + \beta}
{\sqrt{(z^* \beta \beta_\chi +1)^2 - (1- \beta_\chi^2)(1- \beta^2)}},  
\label{eq:boost}
\end{equation}
where $z=\cos\theta$ with $\theta$ being the polar angle 
of $\chi$ (or $\bar \chi$) in the lab frame, 
$z^*=\cos\theta^*$ denotes the $z$ value in the c.m.\ frame, 
$\beta=3/11$,  $\beta_\chi = \sqrt{1-4m_\chi^2/s}$ 
with $m_\chi$ being the mass of DM 
and $s$ being the square of the center-of-mass 
energy of Belle II.
The integration in Eq.~\eqref{eq:di-chi} 
is performed such that both DM particles   
are within the barrel region. 
To do so, we multiply the integrand of Eq.~\eqref{eq:di-chi} 
with a Boolean function which takes the value of unity 
when both $f(z_\chi^*)$ and $f(-z_\chi^*)$ 
take a value between $\cos (128.7^\circ)$ and $\cos (32.2^\circ)$. 
In the massless limit, this leads to 
$ -0.745 < z_\chi^* < 0.745$.

We compute the interaction probability 
between DM ${\chi}$ and the ECL via 
\be
P_\chi = n_E L_E \int d t_{\chi e} \frac{d\sigma_{\chi e}}{d t_{\chi e}}, 
\label{eq:P:chi-e}
\ee
where 
$d \sigma_{\chi e} / dt_{\chi e}$
is the differential cross section for 
the $\chi e^- \to \chi e^-$ process, 
with $t_{\chi e}$ being its $t$-channel Mandelstam variable.  
To compute $P_{\bar\chi}$, one simply substitute $\chi$ with 
$\bar\chi$ in Eq.~\eqref{eq:P:chi-e}. 
The cross sections of 
$d \sigma_{\chi\bar{\chi}}/dz_\chi^*$
and $d \sigma_{\chi e} / dt_{\chi e}$
for different DM models 
are given in appendix \ref{sec:CSequation}.

To analyze the di-cluster events, 
we employ the basic trigger of di-photons, 
since both processes lead to two clusters in the ECL that 
have no preceding tracks in the CDC. 
The basic selection condition of di-photons 
consists of two ECL clusters with $E^* > E^{*}_{\rm th} = 2~\rm GeV$ \cite{Belle-II:2018jsg}, 
where $E^*$ is the energy in the c.m.\ frame. 
For the $\chi e^- \to \chi e^-$ process, 
one has $t_{\chi e} = -2 m_e E_r$ 
where $E_r$ is the electron recoil energy in the lab frame. 
Recoiled electrons then generate EM showers in the ECL, 
leading to clusters with energy $E=E_r$ in the lab frame. 
In the c.m.\ frame, 
the selection condition 
$E^*=\gamma E_r(1-\beta z_\chi)> E_{\rm th}^*$ 
leads to an upper bound on $t_{\chi e}$: 
\be
t_{\chi e}^{\rm max} = \frac{-2 m_e E_{\rm th}^*}{\gamma(1-\beta f(z^*))},
\ee
where $\gamma =1/\sqrt{1-\beta^2}$.
The lower bound on $t_{e\chi}$ is 
\be
t^{\rm min}_{\chi e} = - 
 \left[1-{(m_\chi+m_e)^2 \over s_{\chi e} }\right] \left[ 1-{(m_\chi-m_e)^2 \over s_{\chi e}} \right]  ,      
 \ee
where 
$s_{\chi e } = (p_\chi + p_e)^2
= m_\chi^2 + m_e^2 +m_e \sqrt{s} \gamma (1+\beta \beta_\chi z^*)$.

The dominant background for the DM-induced di-cluster signal 
arises from the SM di-photon process, $e^+e^-\to \gamma\gamma$, 
which in the c.m.\ frame has a production cross section \cite{Peskin:1995ev}
\be
\frac{d\sigma}{dz_\gamma^*}(e^+e^-\to \gamma\gamma)=
\frac{\pi\alpha^2}{s}\frac{1+z_\gamma^{* 2}}{1-z_\gamma^{*2}},
\label{eq:diphoton}
\ee
where $z_\gamma^*=\cos\theta_\gamma^*$ with $\theta_\gamma^*$ 
being the photon polar angle in the c.m.\ frame. 
In order for  both final state photons 
within the barrel region, we integrate $z^*_\gamma$ 
in the range of $-0.745 <z_\gamma^*< 0.745$.
This gives rise to a total di-photon cross section of $\sim 1.4$ nb, 
leading to 
$\simeq 5.1 \times 10^{8}$ ($\simeq 7.0 \times 10^{10}$) 
di-photon events in the SM  
with 362 fb$^{-1}$ (50 ab$^{-1}$) luminosity.

Unlike photons, which deposit nearly their entire energy 
in the ECL, DM only deposits a fraction of its energy in the ECL, 
if the DM mass is not small 
or the DM-electron cross section is not 
strong enough. 
In these regions of the parameter space, 
a detector cut on the energy deposited in the ECL 
could be instrumental in discriminating the signal process 
from the $e^+ e^-\to \gamma \gamma$ process. 
However, because our analysis spans a broad parameter space, where the DM mass ranges from MeV to several GeV and the DM-electron cross-section varies from moderate to very strong, 
we do not 
impose such a detector cut on the energy deposited in the ECL 
in our current study. 
We leave this to a future analysis where one can optimize the 
detector cuts for different regions of the parameter space.

As discussed in the beginning of this section, 
the experimental reconstruction of 
the back-to-back signature in the c.m.\ frame 
from the two DM clusters 
is affected by the partial energy deposition 
in the ECL. 
This is because one cannot 
obtain the true momentum of the DM particle 
in the c.m.\ frame without the precise measurement of 
its energy in the lab frame.
Nevertheless, one can still use the back-to-back signature 
in the transverse plane for the two DM clusters. 
Therefore, the $e^+e^-\to\gamma\gamma\gamma$ events 
are also potential background events, 
if one of the final state photons escapes from the beam direction 
with small transverse momentum.

To assess the background from the 
$e^+e^-\to\gamma\gamma\gamma$ process, 
we generate $10^5$ events  
by using MadGraph with a 1 MeV cut for each photon;  
the total cross section is $\sim 8.6 $ nb. 
We further select events with the following cuts:
(1) There is one photon that either  
escapes in the beam direction or 
has an energy below 0.1 GeV.
(2) The other two photons must 
both have $E^* > E_{\rm th}^*$ and are 
located in the barrel region of the ECL 
with an opening angle larger than $150^\circ$ 
in the c.m.\ frame.  
We find that 5248 events satisfy this selection,
leading to a di-photon cross section of $\sim 0.45$ nb. 
Thus, in the $e^+e^-\to\gamma\gamma\gamma$ process, 
one expects $\sim 1.6\times 10^8$ ($\sim 2.25\times 10^{10}$) 
di-photon events 
for the integrated luminosity of 362 fb$^{-1}$ (50 ab$^{-1}$).

\section{Mono-photon signal at Belle II}
\label{sec:mono-photon}

The mono-photon signature is 
employed as a powerful tool for investigating dark sector 
particles at Belle II \cite{Liang:2019zkb, Liang:2021kgw, Duerr:2019dmv, Duerr:2020muu, Mohlabeng:2019vrz, Kang:2021oes} 
through the $e^+e^-\to\chi\bar{\chi}\gamma$ process.
In the DM-induced mono-photon studies, 
DM is assumed to be invisible 
due to its extremely weak interaction with detectors. 
However, as the interaction increases, the assumption of null-interaction with detectors will eventually fail, 
leading 
to an exclusion region with both an upper boundary (ceiling) and a lower boundary, 
in the $\sigma-m_\chi$ plane, 
where $\sigma$ is the interaction cross section between DM and electrons. 
Below the lower boundary, the interaction strength is too small to yield detectable events; above the upper boundary, DM can leave a visible signal in detectors, 
rendering the mono-photon constraints invalid. 
In this section, we compute the exclusion region 
of the mono-photon signature at Belle II, by taking into account 
the interaction between DM and detectors.

The mono-photon signature at Belle II 
is analyzed first with a set of basic event selection cuts \cite{Belle-II:2018jsg}, 
including an ECL cluster with  $E^{*} > 1.8~\rm GeV$ caused by the photon, 
and  
vetoes
on activities in the following three detectors:
\begin{itemize}
    \item CDC veto: no tracks with $p_T^{*} > 0.2 ~\rm GeV$;
    
    \item ECL veto: no other clusters with $E^{*} > 0.1 ~\rm GeV$; 
    
    \item KLM veto: no KLM clusters outside 
of the $25^\circ$ (3D, COM) cone of the signal photon. 
\footnote{We did not consider this angular cut because, 
in the low-mass region, the photon energy is high, 
and the phase space where the photon and one DM are along the 
same direction is small.}
\end{itemize}

In addition to the basic event selection cuts, 
more advanced detector cuts are employed 
to further reduce the SM backgrounds. 
To compute the exclusion region of the mono-photon signature, 
we adopt the low-mass signal region \cite{Belle-II:2018jsg}.
In our analysis, 
we use the fitting function for the low-mas region \cite{Duerr:2019dmv}: 
\bea
\label{eq:SR1}
\theta_{\mathrm{min}} &=& 5.399^{\circ} 
x^{2} -58.82^{\circ} x +195.71^{\circ}, \\
\theta_{\mathrm{max}}  &=& -7.982^{\circ} x^{2} 
+87.77^{\circ} x -120.6^{\circ},  
\label{eq:lowmass}
\eea 
where $x \equiv E_\gamma^* / \mathrm{GeV}$ with 
$E_\gamma^*$ being the photon energy in the c.m.\ frame, 
$\theta_{\rm min}$ 
and $\theta_{\rm max}$ 
are the minimum and maximum 
angles for the photon in the 
lab frame, namely 
$\theta_{\min }< \theta_\gamma< \theta_{\max }$. 
About 300 mono-photon events from the 
SM backgrounds are 
expected in the low-mass region 
with 20 $\rm fb^{-1}$ of data \cite{Belle-II:2018jsg}. 
We rescale this number  
and obtain 
$\sim$5430 ($\sim$$7.5\times 10^5$) mono-photon background events 
with 362 fb$^{-1}$ (50 ab$^{-1}$) of data.

We next discuss the mono-photon events arising from the DM process. 
The vetoes on the activities in the detectors are normally satisfied if 
the interaction cross section between DM and electrons is small.   
However, as the cross section increases, 
DM can lead to activities in the detectors, 
which can then be vetoed. 
Thus, to compute the upper boundary of the exclusion 
region for the mono-photon signature, one has to 
take into account the three vetoes in the detectors.

We compute the mono-photon events in the low-mass region 
by using 
\be
N_s=\mathcal{L}\int d E_{\gamma}^*  d z_{\gamma}^*  
\frac{d \sigma_{ \chi\bar{\chi}\gamma}}{d E_{\gamma}^*  d z_{\gamma}^*  }  
P_0 ({E}_\chi)^2, 
\label{eq:MP-Sig}
\ee
where $d \sigma_{\chi\bar{\chi}\gamma}/d E_{\gamma}^*  d z_{\gamma}^*$ 
is the mono-photon differential cross section for the 
$e^+ e^- \to \chi \bar{\chi} \gamma$ process, and 
$P_0$ is the probability of DM not inducing significant 
detector activities that are vetoed by the mono-photon detector cuts. 
For the mono-photon signal events in the low-mass region, 
the two final-state DM particles are typically within the ECL coverage. 
We need to compute the interaction probability 
between the two DM particles and the detectors.

In our analysis, we assume that DM is neutral and does not lead to 
any activity in the CDC. 
Thus, we only need to consider the probability of DM-induced activities in ECL and KLM. 
To assess $P_0$, we compute the probability for DM to traverse  
both ECL and KLM without significant activity as follows 
\be
P_0 ({E}_\chi)
= \exp \left[ - \sigma_{\chi e}^v({E}_\chi) 
(n_E \, L_E + n_K \, L_K) \right], 
\label{eq:Ppunch}
\ee 
where $n_K$ is the electron number density of the KLM, 
$L_K$ is the length of the KLM, 
and $\sigma_{\chi e}^v ({E}_\chi)$ is the $\chi-e$ scattering cross section. 
In our analysis, we evaluate $\sigma_{\chi e}^v ({E}_\chi)$ at the 
average energy ${E}_\chi=(\sqrt{s}-E_\gamma^*)/2$. 
To approximate the production 
rate of clusters with energy $>0.1$ GeV 
in ECL and KLM, 
we obtain 
$\sigma_{\chi e}^v$ 
by integrating the differential cross section 
with the recoil energy larger than 0.1 GeV.

\section{mono-cluster signal at Belle II}
\label{sec:monocluster}

In this section, we discuss another possible signature 
due to DM interactions with detectors: the mono-cluster signature.
The mono-cluster signature typically occurs with a 
moderately strong DM-electron interaction cross section 
so that in the $e^+ e^- \to \chi \bar{\chi}$ process, 
only one DM particle leads to an ECL cluster, 
while the other DM penetrates both ECL and KLM without 
any trace.

The DM-induced mono-cluster events 
share a detector response closely resembling that of the mono-photon events. 
In both cases, there is a trackless cluster with a substantial 
amount of energy deposition in the ECL. 
Thus, in the mono-cluster analysis, 
we adopt the mono-photon trigger,  
the basic event selection criteria of the mono-photon channel, 
and the low-mass signal region.

We compute the number of mono-cluster events in the 
low-mass region via
\be
N_s=2 \mathcal{L}   \int dz_{\chi}^*   \frac{d\sigma_{\chi\bar{\chi}}}{dz_{\chi}^*}  
 P_\chi(z_{\chi}^*) P_0(E_{\bar{\chi}}),
\ee
where  
$P_\chi$ is the DM interaction probability with the ECL, 
as given in Eq.~\eqref{eq:P:chi-e}, 
and 
$P_0$ is the DM punch-through probability, 
as given in Eq.~(\ref{eq:Ppunch}). 
Here $E_{\bar{\chi}}$ is the energy of $\bar{\chi}$ in the lab frame, 
which is related to $z^*$ via
$E_{\bar{\chi}} = \sqrt{s}\gamma (1-\beta \beta_\chi z^*)/2$.
The factor of 2 accounts for
the fact that each DM can deposit energy leading to an ECL cluster.

\section{dark matter models}
\label{sec:model}

The DM-induced di-cluster and mono-cluster events 
at Belle II depend on both the DM production process 
and the DM-electron scattering process, as illustrated by 
the two diagrams in Fig.~(\ref{fig:DM:diagram}). 
In contrast, 
DMDD experiments usually only probe  
the DM-electron scattering process. 
This prevents one from making a model-independent 
comparison between the Belle II constraints on 
strongly-interacting DM with DMDD experiments. 
Instead, one has to consider concrete DM models 
to compare constraints from these experiments.
Because Belle II is an electron-positron collider, 
we focus on electro-philic DM models, 
where DM only interacts with electrons in the 
SM sector.

To generate substantial 
DM-induced di-cluster and mono-cluster signal events 
at Belle II, 
a large DM-electron interaction cross section is needed. 
Since DM is neutral, we take the typical interaction cross 
section of a nucleus, $\sim 1$ barn, 
as the benchmark for strongly-interesting DM.
\footnote{Note that the cross section at the level of 
1 barn typically leads to a ceiling at the DMDD experiments; 
see Fig.~(\ref{fig:CSeSIMP}). 
Thus, it is the interaction cross section that 
we aim to study here.}
Such a large DM-SM interaction cross section can be 
realized in models where there is a rather light mediator 
connecting DM and SM particles 
\cite{Bai:2011wy, Daci:2015hca,Li:2014vza, Knapen:2017xzo,Bickendorf:2022buy, Bardhan:2022bdg, Bell:2023sdq}. 
See also Ref.~\cite{Elor:2021swj} 
where the mediator mass at present is significantly reduced 
compared to early universe by a 
phase transition in the dark sector.
\footnote{Note that DM with a large 
self-interacting cross section  
may explain the small scale puzzles \cite{Tulin:2017ara}; 
viable models with sizable DM self-interaction 
cross sections include composite DM 
\cite{Cline:2012is,Cline:2013pca,Cline:2021itd,Cline:2013zca,Kribs:2016cew,Kamada:2022zwb}, 
and models with light mediators \cite{Tulin:2013teo}.}

In our analysis we focus on DM models with light mediators. 
To gauge the required mass range of the mediators 
for a substantial cross section, 
we consider a simple model where a mediator 
couples to DM and electron. 
For non-relativistic DM,  
the DM-electron cross section in DMDD experiments is 
\be
{\sigma}_{\chi e} \sim  
\frac{g_e^2 g_{\chi}^2 \mu^2_{\chi e}}
{(q^2 - m^2)^2}, 
\ee
where 
$q$ ($m$) is the momentum (mass) of the mediator, 
$\mu_{\chi e} = m_\chi m_e/(m_\chi + m_e)$ 
with 
$m_\chi$ ($m_e$) being DM (electron) mass, 
and $g_\chi$ ($g_e$) is the coupling between the mediator and 
DM (electron). 
In the case where $g_e \simeq g_\chi \simeq 1$,  
$m_\chi \gtrsim m_e$ 
and $q^2 \ll m^2$, 
we find that the mediator mass has to be as low as $\sim 1$ MeV 
to yield ${\sigma}_{\chi e} \simeq 1$ b.

Thus, we focus on electro-philic DM models, 
where DM interacts with electrons via a mediator, 
which can be either spin-one or spin-zero. 
We consider the following interaction Lagrangian:
\begin{align}
\mathcal{L}_{\rm int}^V &= Z_{\mu}^{\prime} (g_{\chi}^V \bar{\chi} \gamma^{\mu}\chi+g_e^V \bar{e} \gamma^{\mu}e), 
\label{eq:L:V}
\\
\mathcal{L}_{\rm int}^A &= Z_{\mu}^{\prime} (g_{\chi}^A \bar{\chi} \gamma^5 \gamma^{\mu}\chi+g_e^A \bar{e} \gamma^5 \gamma^{\mu}e), 
\label{eq:L:A}
\\
\mathcal{L}_{\rm int}^S &= \phi (g_{\chi}^S \bar{\chi} \chi + g_e^S  \bar{e} e),
\label{eq:L:S}
\\
\mathcal{L}_{\rm int}^P &= \phi (i g_{\chi}^P \bar{\chi} \gamma^5\chi + i g_e^P  \bar{e}\gamma^5 e),
\label{eq:L:P}
\end{align}
where $Z_{\mu}^{\prime}$ and $\phi$ are 
the spin-one and spin-zero mediators, respectively. 
Here we consider four different cases: 
vector (V), 
axial-vector (A), 
scalar (S), 
and pseudo-scalar (P). 
Here, $g_\chi^i$ ($g_e^i$) is 
the coupling to DM (electron), 
where the superscript $i$ denotes  
the four different mediators.

\section{Constraints on light mediators}
\label{sec:otherexp}

In this section 
we discuss experimental 
constraints on light mediators that 
couple to electrons. 
The best experimental 
constraints currently come from  
the electron $g-2$ measurement, 
electron beam dump experiments,
electron-positron colliders, 
and {M\o ller} scattering.\footnote{See 
Ref.~\cite{Bell:2023sdq} for a recent study 
that compares experimental constraints 
on new light mediators connecting DM 
and hadrons with DMDD constraints.}

\subsection{electron $g-2$}

\begin{figure}[htbp]
\includegraphics[width=0.35 \textwidth]{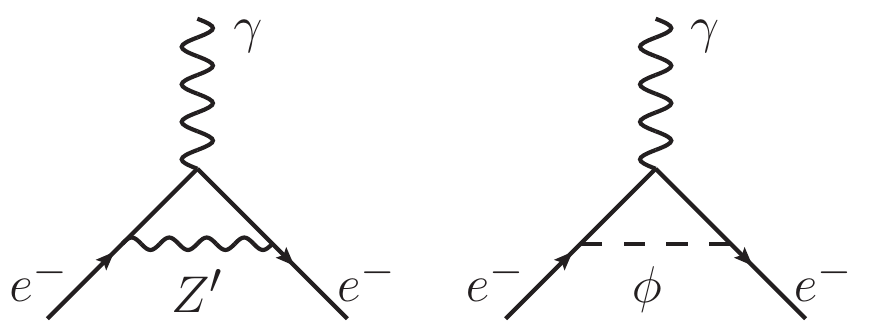}
\caption{One-loop contribution to electron $g-2$ 
from a spin-1 mediator (left) and a spin-0 mediator (right).}
\label{fig:g-2-loop}
\end{figure}

\begin{figure*}[htbp]
\begin{centering}
\includegraphics[width=0.45 \textwidth]{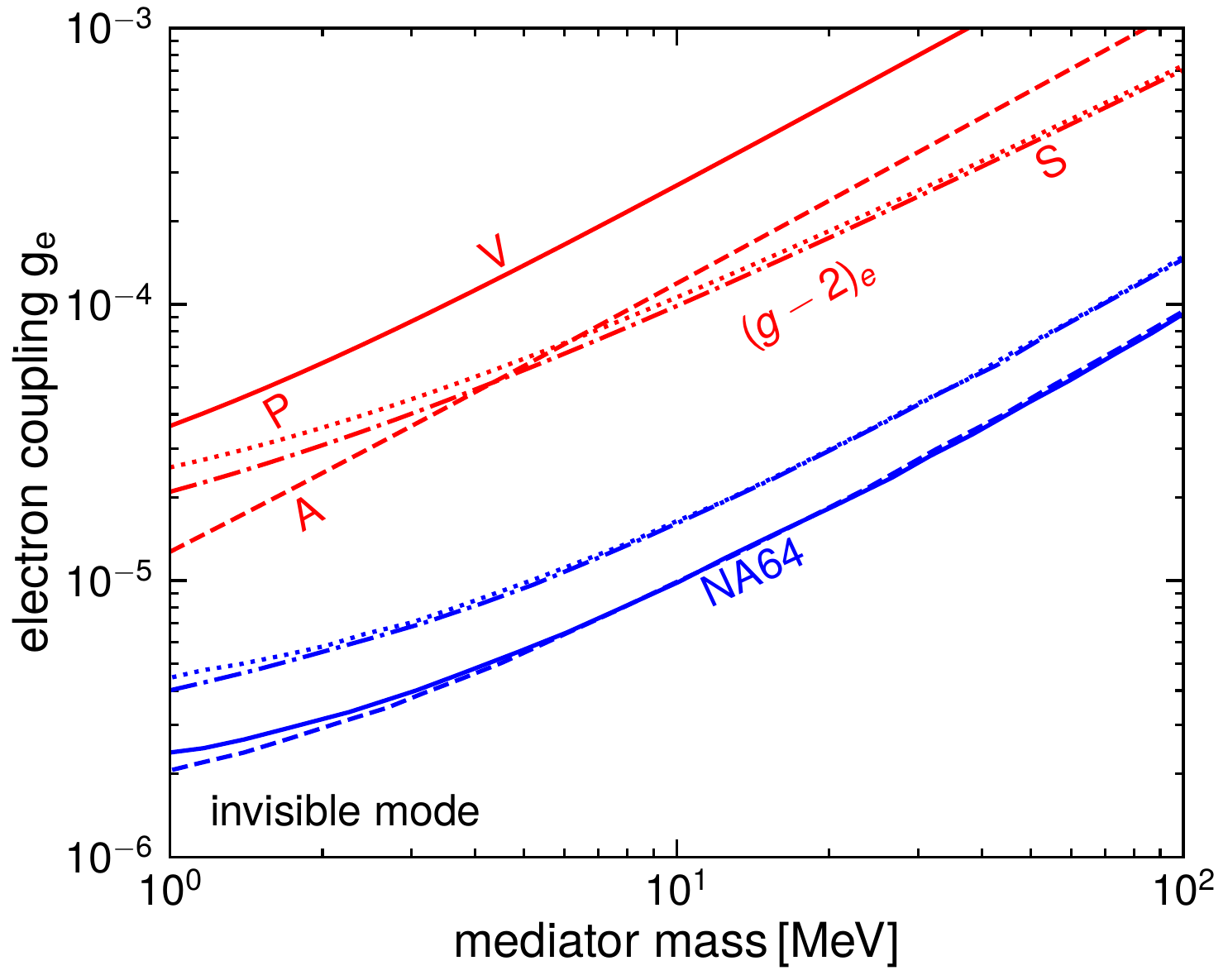}
\includegraphics[width=0.45 \textwidth]{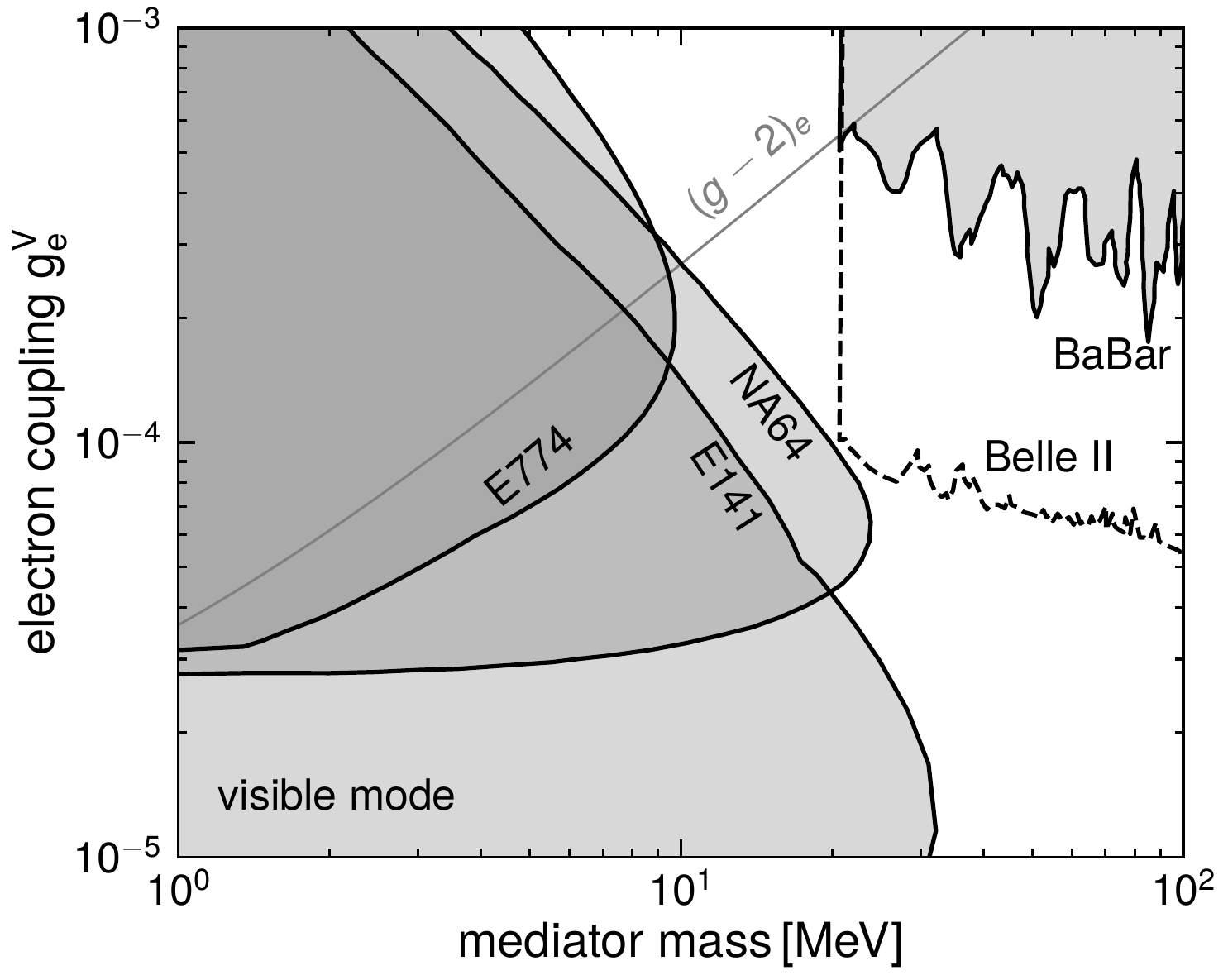}
\caption{{\bf Left}: Constraints on 
mediators that have an electron coupling $g_e$ 
and decay dominantly into DM (invisible mode): 
electron $g-2$ constraints (red)
and NA64 constraints (blue) \cite{ NA64:2021xzo}. 
Four different mediators are shown: 
vector (solid), 
axial-vector (dashed), 
scalar (dot-dashed), 
and pseudo-scalar (dotted). 
{\bf Right}: 
Constraints on the spin-1 
mediator that has a vector coupling to electrons $g_e^V$ 
and decays dominantly into $e^+e^-$ (visible mode). 
The electron $g-2$ constraint is shown as the gray line. 
Other constraints are shown as shaded regions: 
NA64 \cite{NA64:2018lsq}, 
E774 \cite{Bross:1989mp}, 
E141 \cite{Riordan:1987aw}, 
and BaBar \cite{BaBar:2014zli}; 
these constraints are obtained by re-scaling 
the dark photon constraints \cite{Fabbrichesi:2020wbt}
via $g_e=\epsilon e$ with $e$ being the QED coupling and 
$\epsilon$ being the mixing parameter of the dark photon. 
Here we also show the Belle II sensitivity (dashed) with 50 ab$^{-1}$ of data, 
which is obtained by re-scaling 
the dark photon constraints in Refs.~\cite{Belle-II:2018jsg,Ferber:2015jzj}.}
\label{fig:g-2cp}
\end{centering}
\end{figure*}

Light mediators that 
couple with electrons can lead to 
a new physics (NP) contribution 
to the electron $g-2$, as shown in Fig.~(\ref{fig:g-2-loop}).
The contributions to the 
electron $g-2$ from the mediators 
in Eqs.~(\ref{eq:L:V}-\ref{eq:L:P}) 
are 
\cite{Lautrup:1971jf,Leveille:1977rc,Jegerlehner:2009ry} 
\be
\Delta a_e^{\rm NP}=
\frac{g_e^2 \lambda^2}{8 \pi^2} 
\int_0^1 dx 
\frac{Q(x)}{(1-x)(1-\lambda^2x)+\lambda^2x}, 
\label{eq:g-2}
\ee
where 
$\lambda=m_e/m$ with 
$m_e$ ($m$) being the electron (mediator) mass,
$g_e$ denotes the various couplings to electrons 
in Eqs.~(\ref{eq:L:V}-\ref{eq:L:P}), 
and $Q(x)$ are  
\begin{align}
Q_V &=2x^2 (1-x),\\
Q_A &=2x(1-x)(x-4)-4\lambda^2x^3,\\
Q_S &=x^2(2-x),\\
Q_P &=-x^3,  
\end{align}
where the subscript denotes the four types of mediators in 
Eqs.~(\ref{eq:L:V}-\ref{eq:L:P}).

The interpretation of electron $g-2$ data 
depends on the experimental determination of 
the fine structure constant $\alpha$. 
By using the $\alpha$ value measured with  
rubidium (Rb) atoms \cite{Morel:2020dww} 
and cesium (Cs) atoms \cite{Parker:2018vye}, 
it is shown in Ref.~\cite{Davoudiasl:2023huk} that 
the new electron $g-2$ measurement \cite{Fan:2022eto} 
has a 2.2 $\sigma$ and -3.7 $\sigma$ deviations 
from the SM prediction 
\cite{Aoyama:2017uqe}:
\bea
\Delta a_e({\rm Rb})&=& (34\pm 16) \times 10^{-14},
\label{eq:Dg-21}\\
\Delta a_e({\rm Cs})&=& (-101\pm 27) \times 10^{-14}. 
\label{eq:Dg-22}
\eea
Given the intricate aspects of this measurement, 
we adopt a cautious approach in constraining new physics models: 
We add a $2\sigma$ to the central deviations 
in Eqs.~(\ref{eq:Dg-21}-\ref{eq:Dg-22}) and then 
use the largest deviation to constrain  
new physics contributions  
regardless the sign. 
Thus, the new physics contributions should satisfy 
\be
|\Delta a_e^{\rm NP}| \lesssim 155 \times 10^{-14}. 
\ee
Fig.~(\ref{fig:g-2cp}) shows the 
constraints on the four types of mediators in 
Eqs.~(\ref{eq:L:V}-\ref{eq:L:P}).

\subsection{electron beam dump experiments and BaBar }

Light mediators that couple to electrons 
can be searched for both in electron beam dump experiments 
and in electron-positron colliders. 
Unlike the electron $g-2$ constraint, 
which is insensitive to the interaction between DM 
and the mediator (at least in the leading order), 
the experimental constraints 
from electron beam dump experiments 
and electron colliders are very sensitive to 
the invisible decay width of the mediator. 
Thus, we discuss the constraints in two cases: 
(1) the invisible mode, where the mediator dominantly decays into 
DM;
(2) the visible mode, where the mediator dominantly decays into 
SM particles.

\subsubsection{Invisible mode}

The invisible mode can occur 
in the case where 
$ m > 2 m_\chi$ and $g_\chi \gg g_e$. 
For the mediator mass in the range 
of $\sim(1-100)$ MeV, the leading constraints 
come from the missing momentum search at NA64 
\cite{NA64:2021xzo}, as shown 
in the left panel figure of Fig.~(\ref{fig:g-2cp}). 
The NA64 constraints on the electron coupling $g_e$ 
are about one order of magnitude 
stronger than the electron $g-2$ constraints.

We note that the NA64 constraints \cite{NA64:2021xzo} 
are analyzed under the assumption 
that DM does not interact with detectors, 
thereby resulting in a missing momentum signature. 
However, this assumption 
becomes invalid in the case of strongly-interacting DM,  
where DM exhibits a substantial interaction cross section with SM particles 
and is likely to be absorbed by the NA64 detectors. 
Therefore, the NA64 constraints shown in 
the left panel of Fig.~(\ref{fig:g-2cp}) 
are not applicable to strongly-interacting DM.\footnote{See also Ref.~\cite{Arefyeva:2022eba} 
where the NA64 constraints on millicharged particles 
are investigated by taking into account various effects due to 
significant interactions between millicharged particles 
and NA64 detectors. In particular, 
there is a ceiling on the millicharge, 
above which the NA64 constraints 
are not applicable.}

\subsubsection{Visible mode}

The visible mode can occur in the case where 
$ 2m_\chi > m$. 
For the mediator mass in the range 
of $\sim(1-100)$ MeV, the primary decay channel 
is the $e^+e^-$ final state. 
In this case, the leading constraints 
come from electron beam dump experiments 
and electron-positron colliders.

The right panel figure of Fig.~(\ref{fig:g-2cp}) 
shows the constraints on the visible signals 
arising from 
the spin-1 mediator 
that has a vector coupling to electrons. 
Here, we re-scale the dark photon constraints 
\cite{NA64:2018lsq, BaBar:2014zli}
that arise only from electron couplings; 
the re-scaling is done via 
$g_e = e \epsilon$ where $\epsilon$ is the 
mixing parameter in dark photon models 
\cite{Holdom:1985ag,Foot:1991kb,Kors:2005uz,Feldman:2007wj,Fabbrichesi:2020wbt}. 
As shown in the 
right panel figure of Fig.~(\ref{fig:g-2cp}), 
the leading constraints come from  
BaBar \cite{BaBar:2014zli},
and electron beam dump experiments including 
NA64 \cite{NA64:2021xzo, Banerjee:2019pds}, 
E141 \cite{Riordan:1987aw}, 
and E774 \cite{Bross:1989mp}.\footnote{We note that 
many accelerator constraints on the dark photon 
rely on its production in meson decays: 
for example, 
$\pi^0 \to \gamma A'$, $\eta \to \gamma A'$, and $\Delta \to N A'$ in HADES \cite{HADES:2013nab}, 
$\pi^0, \eta\to\gamma A'$ decays in PHENIX \cite{PHENIX:2014duq}, 
$\pi^0$ decays in NA48 \cite{NA482:2015wmo}, 
and 
$\phi \to \eta A'$ decays in KLOE \cite{KLOE-2:2012lii}. 
Therefore, these constraints are not applicable 
to light mediators that only couple to electrons.}

Remarkably, 
there is a large portion of parameter space
consistent with various constraints: 
${5 ~\rm MeV} \lesssim m \lesssim 100 {~\rm MeV}$ 
and $10^{-5} \lesssim g_e \lesssim 10^{-3}$. 
The beam dump experiments lose sensitivity in this region 
primarily because of the short lifetime of the mediators, 
so that the mediators are likely to decay in the dump. 
The termination of the BaBar constraints at $\sim 20$ MeV 
is largely due to the large SM background in the low 
invariant mass bins of the 
$e^+ e^-$ pair in the process of 
$e^+ e^- \to \gamma Z' \to \gamma e^+ e^-$.

%%%

\subsection{M\o ller scattering}

Because the contributions to the electron $g-2$ from 
the vector and axial-vector couplings have opposite signs 
\footnote{In the leading order, 
there is no interference term 
between $g_e^V$ and $g_e^A$ in the electron $g-2$ calculation; 
see e.g., \cite{Leveille:1977rc} \cite{Bodas:2021fsy}.}, 
the spin-1 mediator can have a vanishing contribution 
to the electron $g-2$ if both vector 
and axial-vector couplings are present. 
However, such a scenario is strongly constrained  
by the {M\o ller} scattering, $e^- e^- \to e^- e^-$, 
from the SLAC E158 experiment \cite{SLACE158:2005uay}.   
For the $Z'$ mass $\lesssim 100$ MeV, 
the E158 constraint is roughly mass-independent: 
$|g_e^Vg_e^A| \lesssim 10^{-8}$ 
\cite{Kahn:2016vjr, Bodas:2021fsy}. 
Thus, it is difficult to have both vector and axial-vector couplings   
sizable simultaneously. 
In fact, in the democratic setting, namely $g_e^V \sim g_e^A$, the 
SLAC E158 experiment imposes a stronger constraint than the electron 
$g-2$ for the mediator mass $\gtrsim 10$ MeV;  
for the mediator mass $\lesssim 3$ MeV, however, 
the electron $g-2$ constraint can be somewhat alleviated 
with $g_e^V \sim g_e^A$.

Note that for a spin-one mediator that couples with electrons, 
parity-violating effects in M\o ller scattering 
arise when both vector and axial-vector couplings are present. 
Thus, in a new physics model with two electro-philic spin-one mediators, 
where one mediator only has a vector coupling with electrons 
and the other mediator only has an axial-vector coupling with electrons,  
constraints from the parity-violating asymmetry measurement in M\o ller scattering are absent. 
Moreover, cancellations to the electron $g-2$ between the two mediators 
also render the electron $g-2$ constraint unimportant, 
if the vector and axial-vector couplings are comparable. 
Therefore, in this case, the couplings between the two mediators 
and electrons can be both sizable.

\begin{figure*}[htbp]
\begin{centering}
\includegraphics[width=0.45 \textwidth]{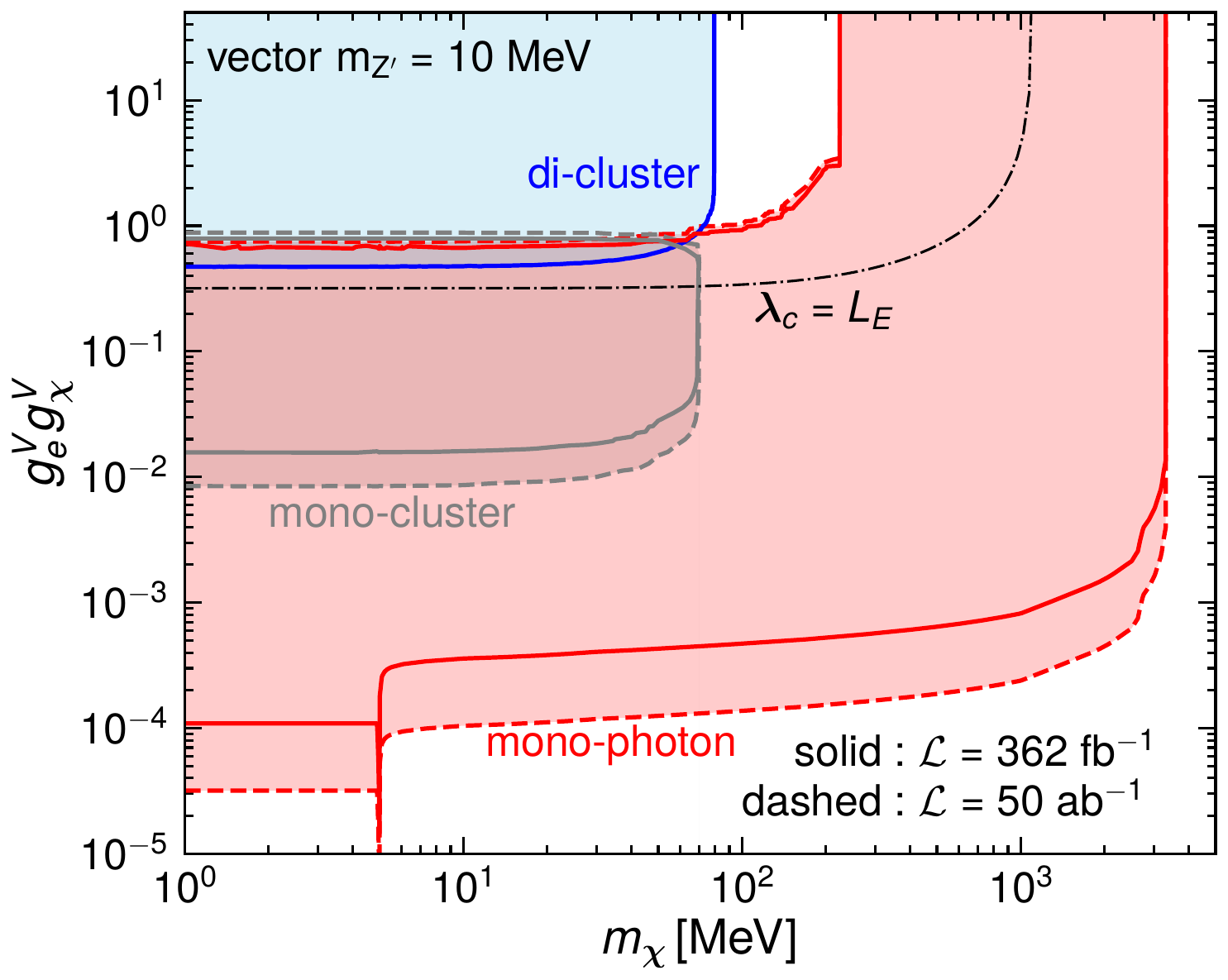}
\includegraphics[width=0.45 \textwidth]{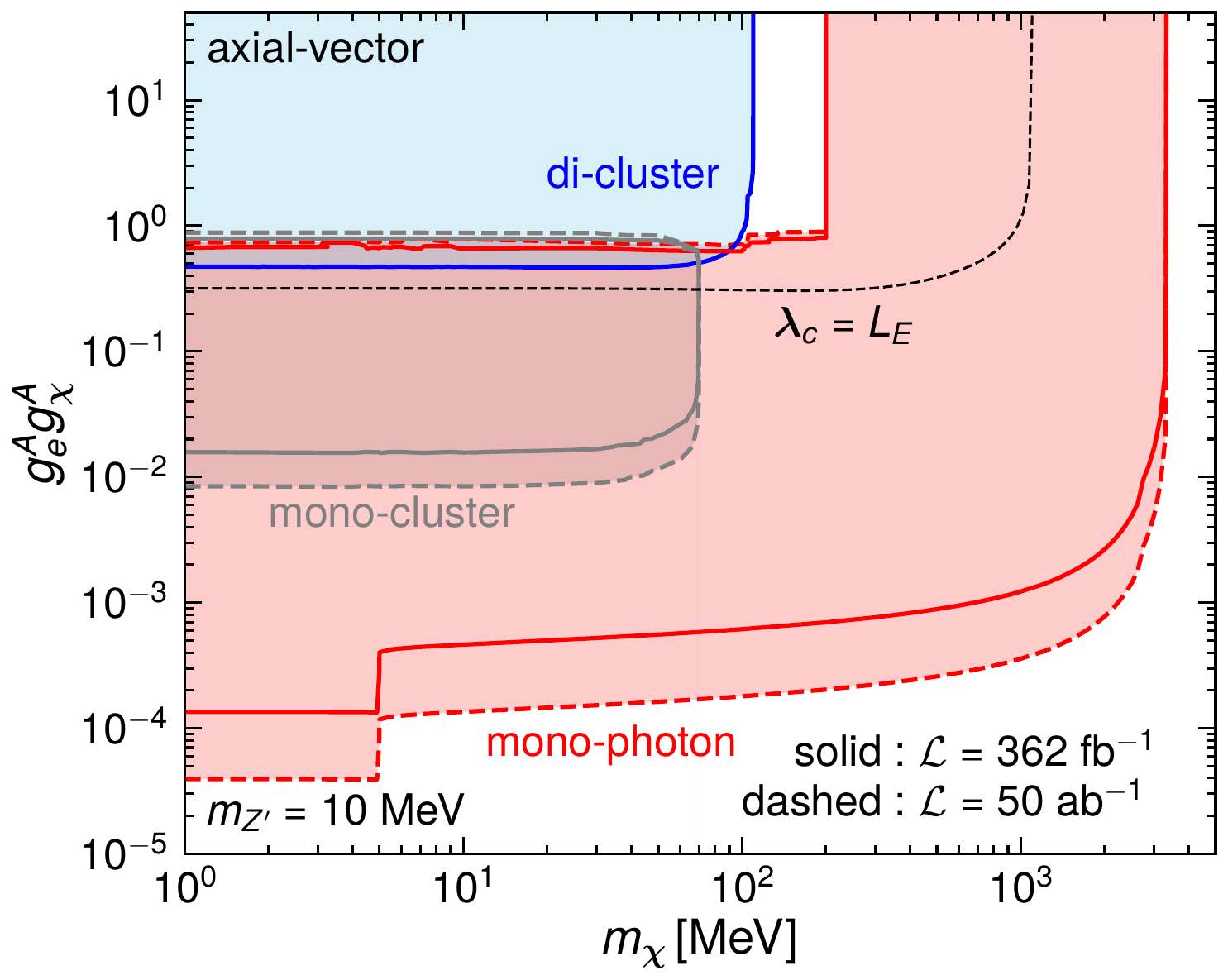}
\includegraphics[width=0.45 \textwidth]{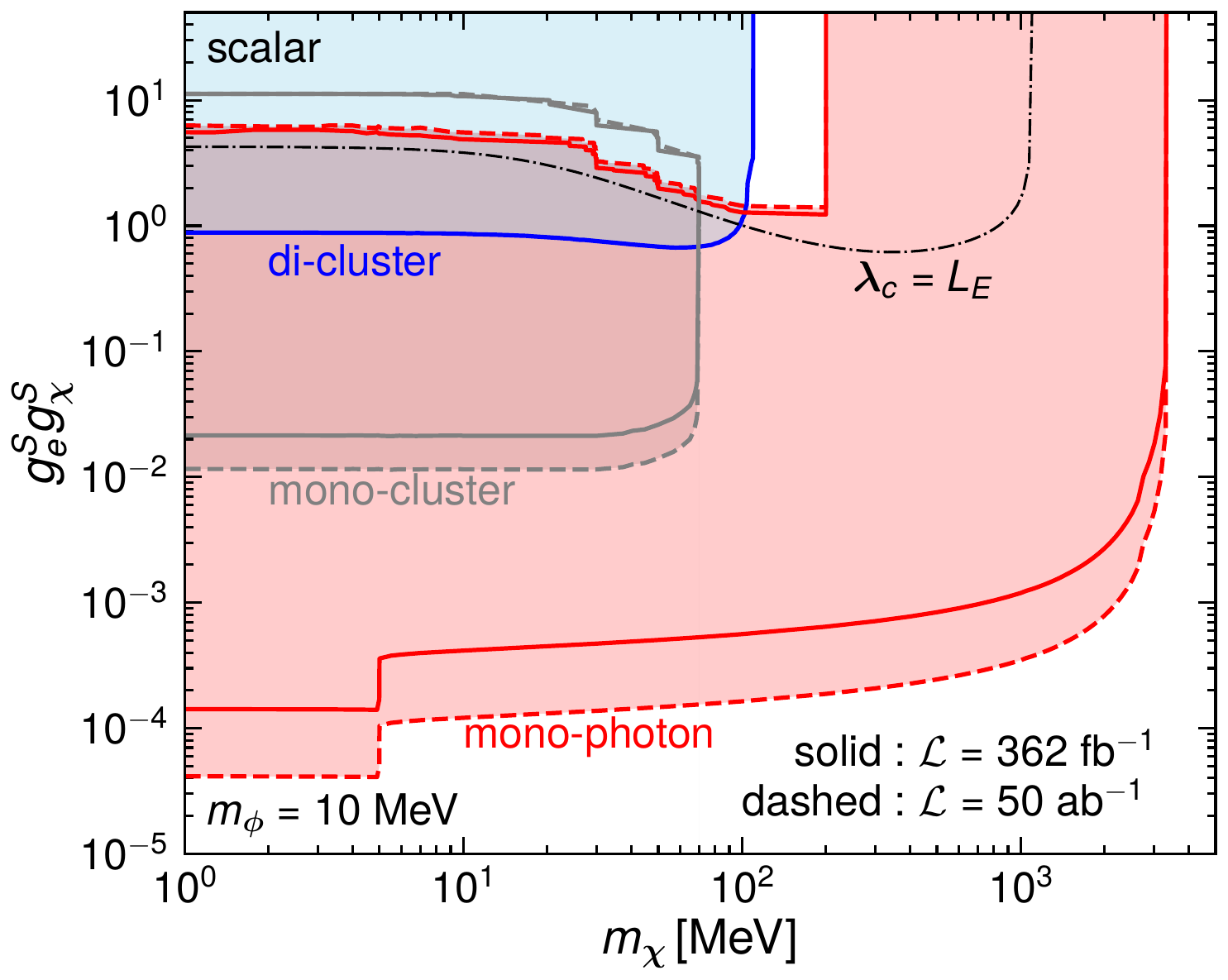}
\includegraphics[width=0.45 \textwidth]{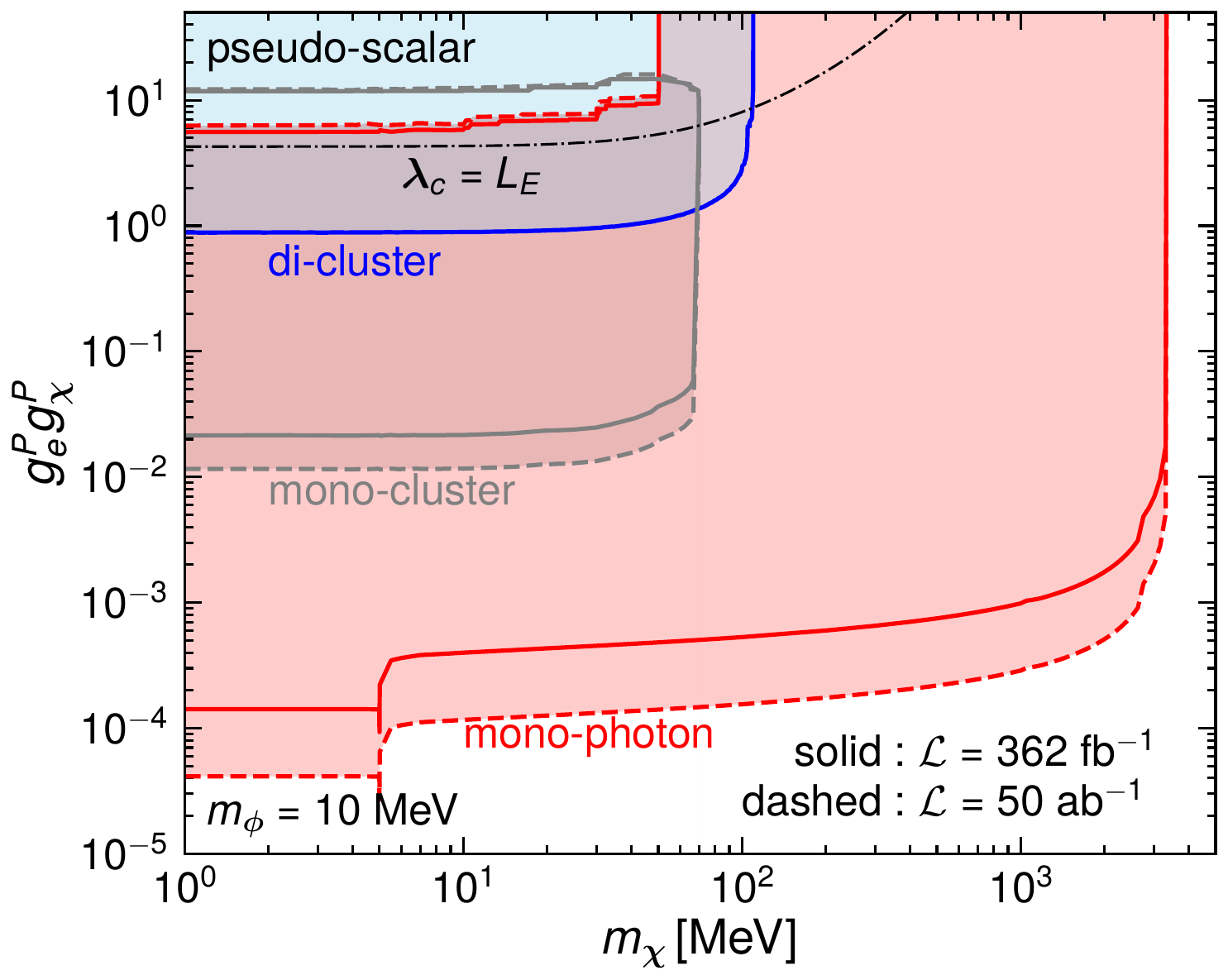}
\caption{Belle II 90\% C.L.\ exclusion regions 
for mediators with a $10$ MeV mass,  
from different channels: 
(1) mono-photon (red), 
(2) mono-cluster (gray),
(3) di-cluster (blue). 
Solid (dashed) lines denote data 
with $362$ fb$^{-1}$ \cite{Bertacchi:2023kdm} ($50$ ab$^{-1}$) luminosity. 
Four different mediators are considered: 
(1) vector (upper-left), 
(2) axial-vector (upper-right),  
(3) scalar (lower-left), 
(4) pseudo-scalar (lower-right).
The black dot-dashed curves denote $\lambda_c=L_E$ where $L_E=37$ cm. 
For the di-cluster channel, the dashed and solid lines merge due to the dominance of the systematic uncertainty over the statistical uncertainty in this channel.
}
\label{fig:CP:10}
\end{centering}
\end{figure*}

\begin{figure*}[htbp]
\begin{centering}
\includegraphics[width=0.45 \textwidth]{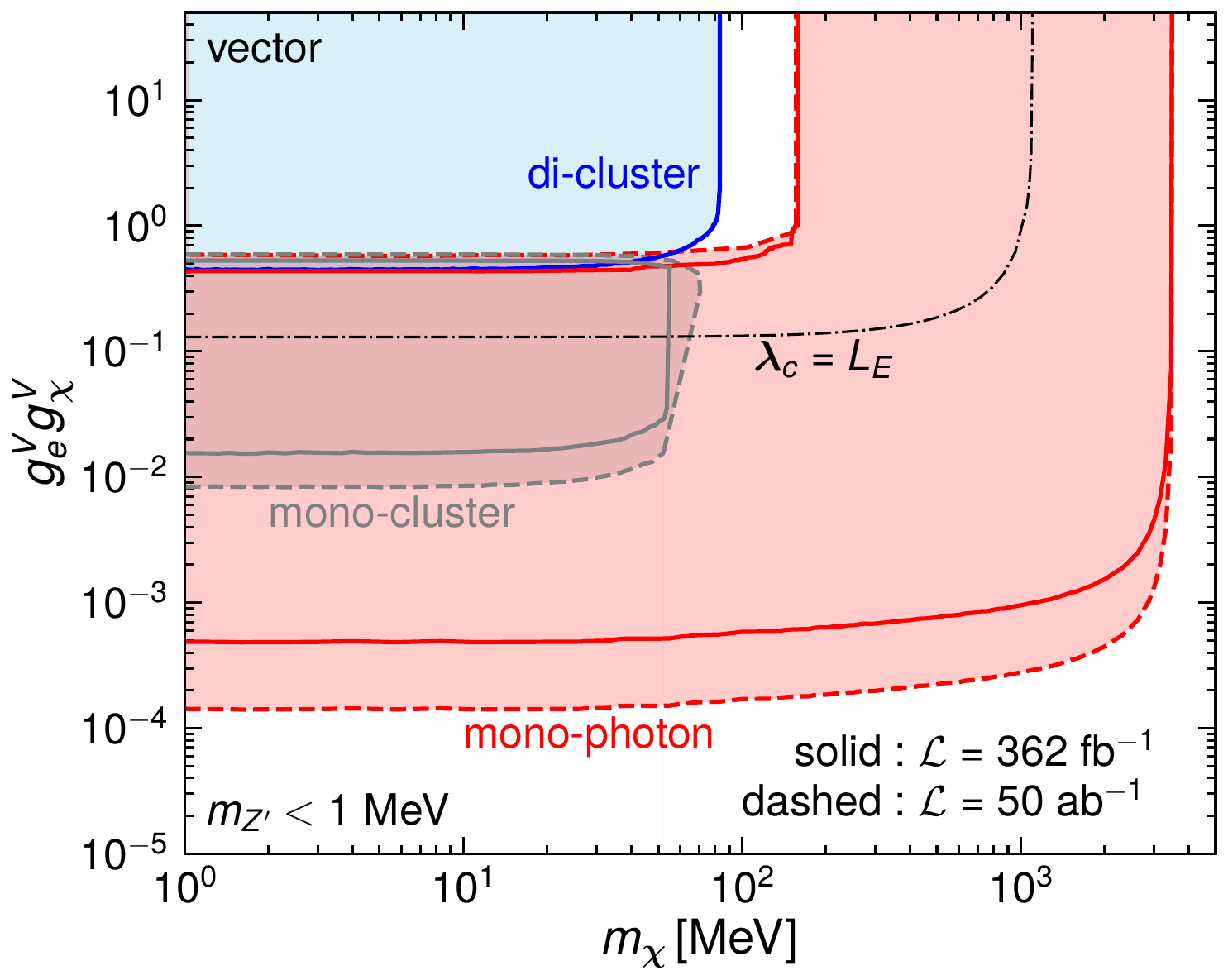}
\includegraphics[width=0.45 \textwidth]{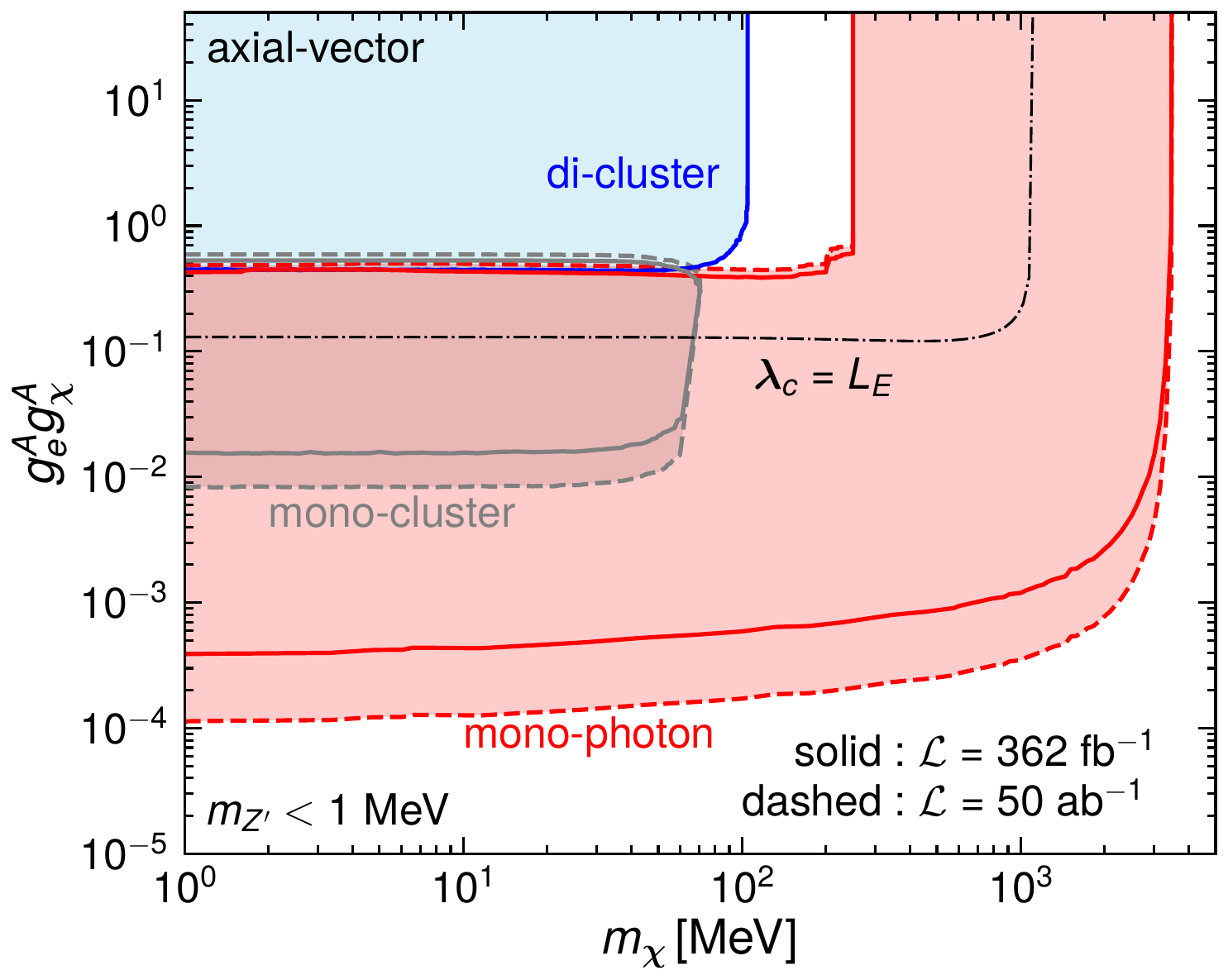}
\includegraphics[width=0.45 \textwidth]{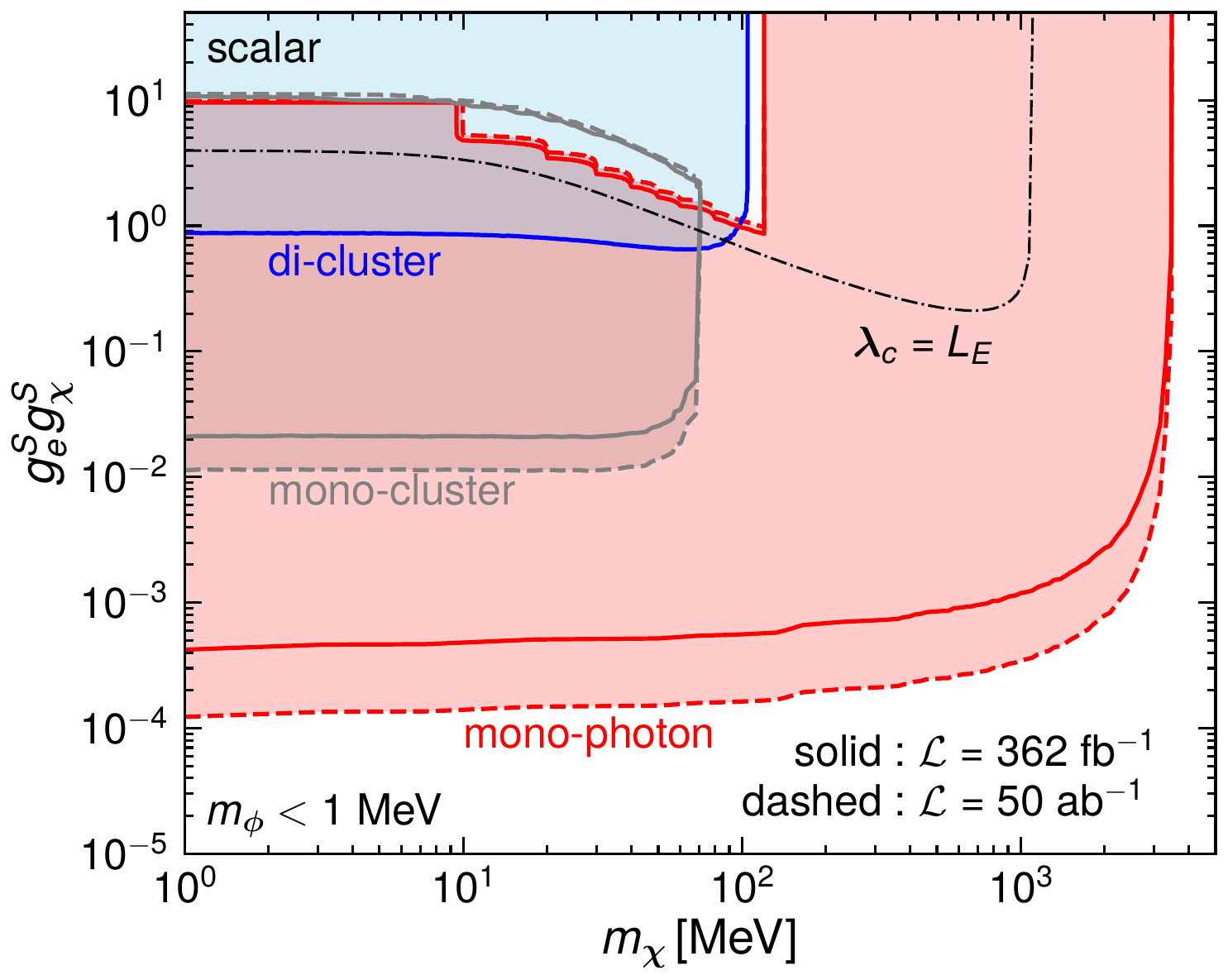}
\includegraphics[width=0.45 \textwidth]{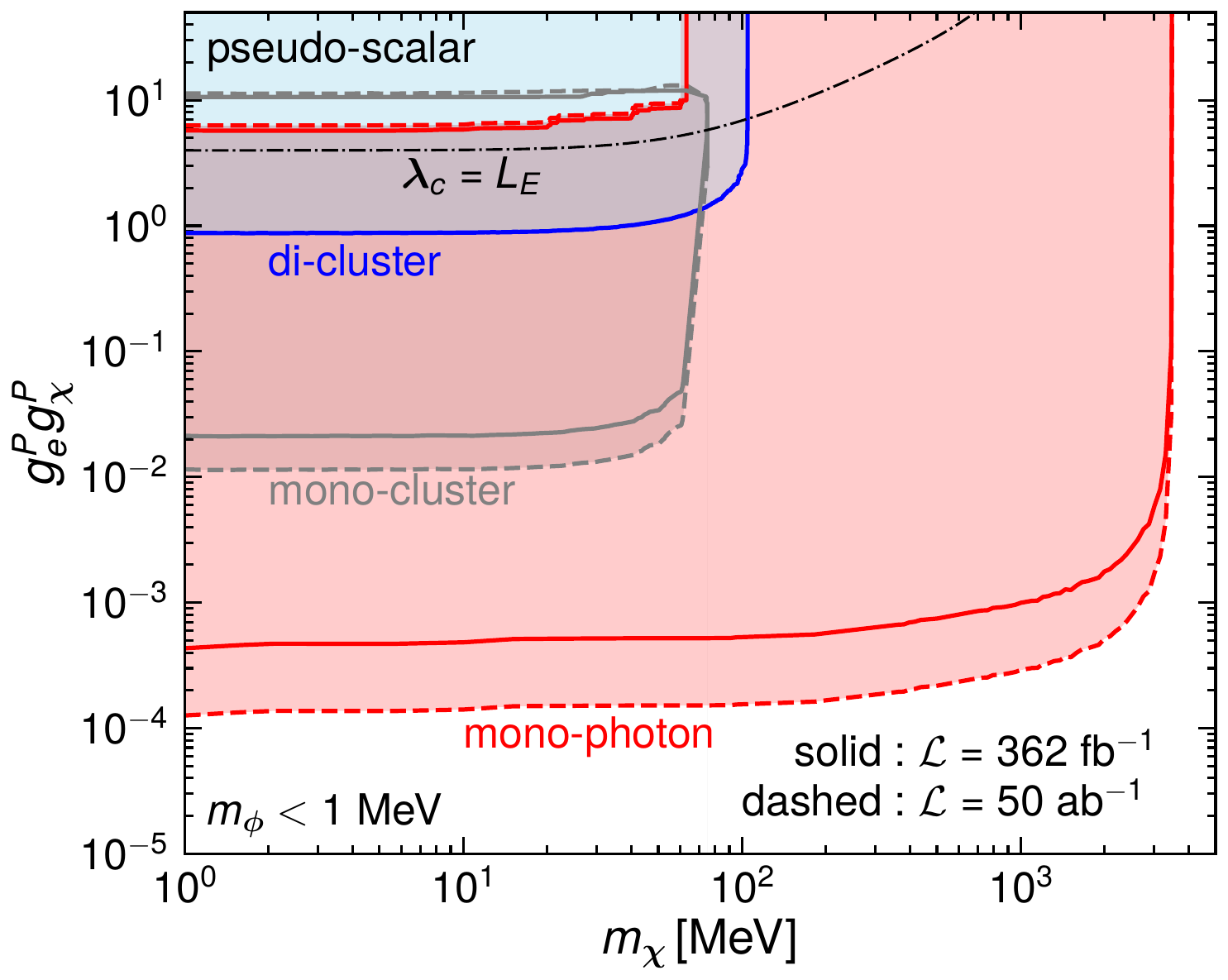}
\caption{Same as Fig.~(\ref{fig:CP:10}), but for the mediator mass $<1$ MeV.}
\label{fig:CP:1}
\end{centering}
\end{figure*}

\section{Belle II sensitivity}
\label{sec:CPBelle2}

In this section we analyze constraints
from three different channels at Belle II: 
the di-cluster, mono-cluster, and mono-photon channels.
We compute the 90\% C.L.\ limits for all the three channels
via the criterion of $\chi^2= N_s^2/\sigma_b^2=2.71$ by assuming the 
Gaussian distribution, 
where $N_s$ is the number of signal events 
and $\sigma_b$ is the uncertainty of the background,  
which includes both the statistical uncertainty 
and the systematic uncertainty. 
For the statistical uncertainty, we use 
$\sigma_b = \sqrt{N_b}$, where $N_b$ is the number of the background events. 
For the systematic uncertainty in the di-cluster channel, 
we adopt $\sigma_b = 1\% N_b$, 
which is the systematic uncertainty of the di-photon events 
\cite{Belle-II:2019usr}. 
For the mono-cluster and mono-photon channels, we have neglected the systematic uncertainty 
associated with mono-photon events in the SM, 
due to the scarcity of literature on such systematic uncertainties 
and the relatively small number of background events.\footnote{If a 
systematic uncertainty of ${\cal O}(1\%)$ were assumed, 
the constraints from the mono-cluster and mono-photon channels  
with 50 ab$^{-1}$ would become more or less the same as 
those obtained from 362 fb$^{-1}$ in 
Figs.~(\ref{fig:CP:10}-\ref{fig:CP:1}).}

Fig.~(\ref{fig:CP:10}) shows the Belle II constraints for 
mediators with a mass of 10 MeV; 
Fig.~(\ref{fig:CP:1}) shows the Belle II constraints for 
mediators with a mass of 1 MeV.  
Because the Belle II constraints presented here 
have a rather weak dependence on the mediator mass 
in the mass range of $\lesssim 1 ~\rm MeV$, 
the constraints given in Fig.~(\ref{fig:CP:1}) 
are applicable to mediators with mass $< 1 ~\rm MeV$.
We compute the constraints  
both with the current data of 
$\mathcal{L}=362$ fb$^{-1}$ \cite{Bertacchi:2023kdm}, 
and with the total expected data of $\mathcal{L}=50$ ab$^{-1}$.

As shown in both Fig.~(\ref{fig:CP:10}) and Fig.~(\ref{fig:CP:1}),
for both the mono-photon and mono-cluster channels, 
the exclusion regions typically exhibit 
both a lower boundary 
and an upper boundary;
the parameter space below the lower boundary 
or above the upper boundary (ceiling) is 
not constrained. 
In contrast, the constraints from the di-cluster channel manifest as exclusion regions with only lower boundaries, namely, without ceilings.

The ceilings in the mono-photon and mono-cluster channels 
arise when DM starts to have a substantial interaction cross section 
with detectors. 
The locations and shapes of these ceilings 
can also be 
estimated via $\lambda_c = L_E$ 
with the energy of the incident DM being 5 GeV, 
where $L_E=37$ cm is the length of the ECL,
as shown in 
Fig.~(\ref{fig:CP:10}) and Fig.~(\ref{fig:CP:1}). 
Significantly below the $\lambda_c = L_E$ curve, 
DM rarely interacts with the ECL, 
manifesting itself as a missing signature; 
significantly above the $\lambda_c = L_E$ curve, 
the multiple-scatterings become important.

The di-cluster constraints with the current data of 
$\mathcal{L}=362$ fb$^{-1}$ \cite{Bertacchi:2023kdm} 
can already probe a significant portion of the 
parameter above the ceilings of both 
the mono-photon channel and the mono-cluster channel.
We find that increasing the current Belle II data to the total expected data of 
$\mathcal{L}=50$ ab$^{-1}$ only improves the lower 
boundaries of the di-cluster constraints, 
which, however, have already been ruled out by 
either the mono-photon channel 
or the mono-cluster channel.

For the pseudo-scalar mediator case,
the di-cluster constraints 
can exclude the entire parameter space 
above the ceilings of the mono-photon and mono-cluster channels. 
However, for the other three mediators, 
there is a significant portion of 
parameter space 
unconstrained by the di-cluster channel, 
above these two ceilings. 
We note that the parameter space 
that is probed by the mono-cluster channel 
predominantly falls within the mono-photon exclusion region; 
in instances where it extends beyond the mono-photon ceiling, 
it is concurrently constrained by the di-cluster channel.

We note that Belle II constraints are insensitive to the mediator mass $m$, 
if $m \ll \sqrt{s}$ where $\sqrt{s} \simeq 10.5$ GeV 
is the colliding energy of Belle II. 
This explains that the Belle II constraints given in 
Fig.~(\ref{fig:CP:10}) and Fig.~(\ref{fig:CP:1}) 
are nearly the same, 
except the lower boundary of the mono-photon exclusion region, 
in the parameter space  
where the mediator can decay into a pair of the DM. 
As shown in Fig.~(\ref{fig:CP:10}) where the mediator mass is 10 MeV, 
the lower boundary of the mono-photon exclusion region 
for $m_\chi < 5$ MeV 
is shifted downward significantly 
compared to that for $m_\chi > 5$ MeV. 
This is because when $m_\chi < 5$ MeV, 
DM can be produced via the resonance of the mediator, 
thus boosting its production cross section.

The mono-photon ceilings only exist  
for small DM mass and 
start to disappear when $m_\chi \gtrsim 0.2~\rm GeV$, 
as shown in Fig.~(\ref{fig:CP:10}) and Fig.~(\ref{fig:CP:1}).
However, this is primarily due to the 
single-scattering assumption in our calculation, 
in which the maximum value of the electron recoil energy 
becomes less than 0.1  GeV, if the DM mass exceeds 0.2 GeV.
As the interaction cross section increases, 
multiple scatterings become important. 
Even if the energy deposited in the ECL in one scattering 
is low, 
the sum of the recoil energies in all DM scatterings with the ECL 
can still be significant, leading to an ECL cluster with energy 
larger than 0.1 GeV. 
Therefore, the ceiling for the case of $m_\chi \gtrsim  {0.2}~\rm GeV$ 
should appear with a sufficiently large interaction cross section. 
\footnote{However, when the DM mass is large such that its kinetic energy 
is below the energy threshold of 0.1 GeV, the ceiling of the 
mono-photon channel cannot appear.}
In our analysis, we have neglected multiple scatterings due to 
the complexity. We leave that to a future study.

Both di-cluster and mono-cluster channels 
lose sensitivity when $m_\chi \gtrsim 0.1 ~\rm GeV$. 
This is also largely due to the 
single-scattering
assumption 
used in our analysis. 
We require the cluster energy to be larger 
than 2 GeV (1.8 GeV)
for the di-cluster (mono-cluster) channel, 
which cannot be realized in a single elastic scattering 
for a DM with $m_\chi \gtrsim 0.1 ~\rm GeV$ and $E_\chi \simeq 5~\rm GeV$.

While the Belle II channels analyzed in this section 
constrain the product of $g_e$ and $g_\chi$, 
the experimental constraints analyzed in section \ref{sec:otherexp} 
only limit $g_e$. 
As shown in Fig.~(\ref{fig:g-2cp}), the electron $g-2$ constraints 
apply to both the invisible and visible modes of the mediators 
and lead to a limit of $g_e \lesssim (10^{-5}-10^{-4})$,  
for the mediator mass $\sim(10-100)$ MeV. 
Combining these two types of constraints, we find that 
the mono-photon ceilings 
or the lower boundaries of 
the di-cluster exclusion regions in 
Fig.~(\ref{fig:CP:10}) and Fig.~(\ref{fig:CP:1})  
are already in the parameter region of  
$g_\chi \sim (10^{3}-10^{4})$ where   
perturbative calculations start to fail. 
However, as discussed in section~\ref{sec:otherexp}, 
couplings to electrons can be quite substantial 
in a NP model with two spin-one mediators, 
where one mediator only has vector couplings
and the other only has axial-vector couplings;
in such a model, 
electron $g-2$ constraints can be mitigated 
by cancellations between the two mediators,  
and NA64 constraints can be alleviated 
by the strong DM-electron interaction cross section. 
To summarize, 
we find that although 
it is challenging to perturbatively analyze the strongly-interacting 
DM signals at Belle II in a simple mediator model, 
there are viable models with more than one mediator.

\subsection{Search for mediators via visible channels}

In addition to the DM processes,  
the light mediators can also be searched for 
at Belle II via the ``visible'' channels. 
Unlike the DM processes,
which probe both $g_\chi$ and $g_e$, 
the visible channels only probe $g_e$. 
Here we estimate the Belle II sensitivities on new light  
spin-one mediators with vector couplings, 
from the $e^+e^-$ final state 
via the following two processes: 
(1) the radiative return process, 
$e^+e^- \to \gamma Z' \to \gamma e^+e^-$, 
in which the mediator $Z'$ is produced 
nearly on-shell  
and subsequently decays into an electron-positron pair; 
(2) the $t$-channel process mediated by $Z'$ 
that contributes to the Bhabha scattering.

The mediator in the radiative return process 
can be probed via the resonance search 
where the invariant mass of the final state electron-positron 
pair is reconstructed. 
The right panel figure of Fig.~(\ref{fig:g-2cp}) 
shows the Belle II sensitivity in this channel 
(with 50 ab$^{-1}$ of data),  
which is obtained by properly re-scaling 
the dark photon constraints 
in Refs.~\cite{Belle-II:2018jsg,Ferber:2015jzj}. 
We note that the expected Belle II limits  
(as well as the BaBar constraints in this channel, 
as also shown in the right panel figure of Fig.~(\ref{fig:g-2cp})) 
only probe mediators with  
mass $\gtrsim 20$ MeV, 
and are not relevant to our current study 
which focuses on 
mediators with mass $\lesssim 10$ MeV.

Measurement of 
Bhabha scattering data can impose constraints on the new physics  
process of $e^+e^- \to e^+e^-$ mediated by a $t$-channel $Z'$. 
The leading new physics contribution arises 
from the interference term between the SM process and the $Z'$ process: $N_s \simeq 2(g_e^V/e)^2 N_{\rm SM}$, 
where $N_{\rm SM}$ is the number of SM background events. 
To compute the 90\% CL limits, 
we use the criterion of $\chi^2= N_s^2/\sigma_b^2=2.71$, 
which is the same as the DM channels. 
We adopt 0.6\%$N_{\rm SM}$ \cite{Belle-II:2019usr} for the systematic uncertainty; 
the statistical uncertainty is negligible due to the 
large number of Bhabha scattering events. 
We find that 
the Belle II sensitivity at 90\% CL from Bhabha scattering is 
$g_e^V \lesssim 0.02$. 
We note that the constraints from Bhabha scattering are significantly weaker 
than those shown in the right panel of Fig.~(\ref{fig:g-2cp}).

\begin{figure*}[htbp]
\begin{centering}
\includegraphics[width=0.45 \textwidth]{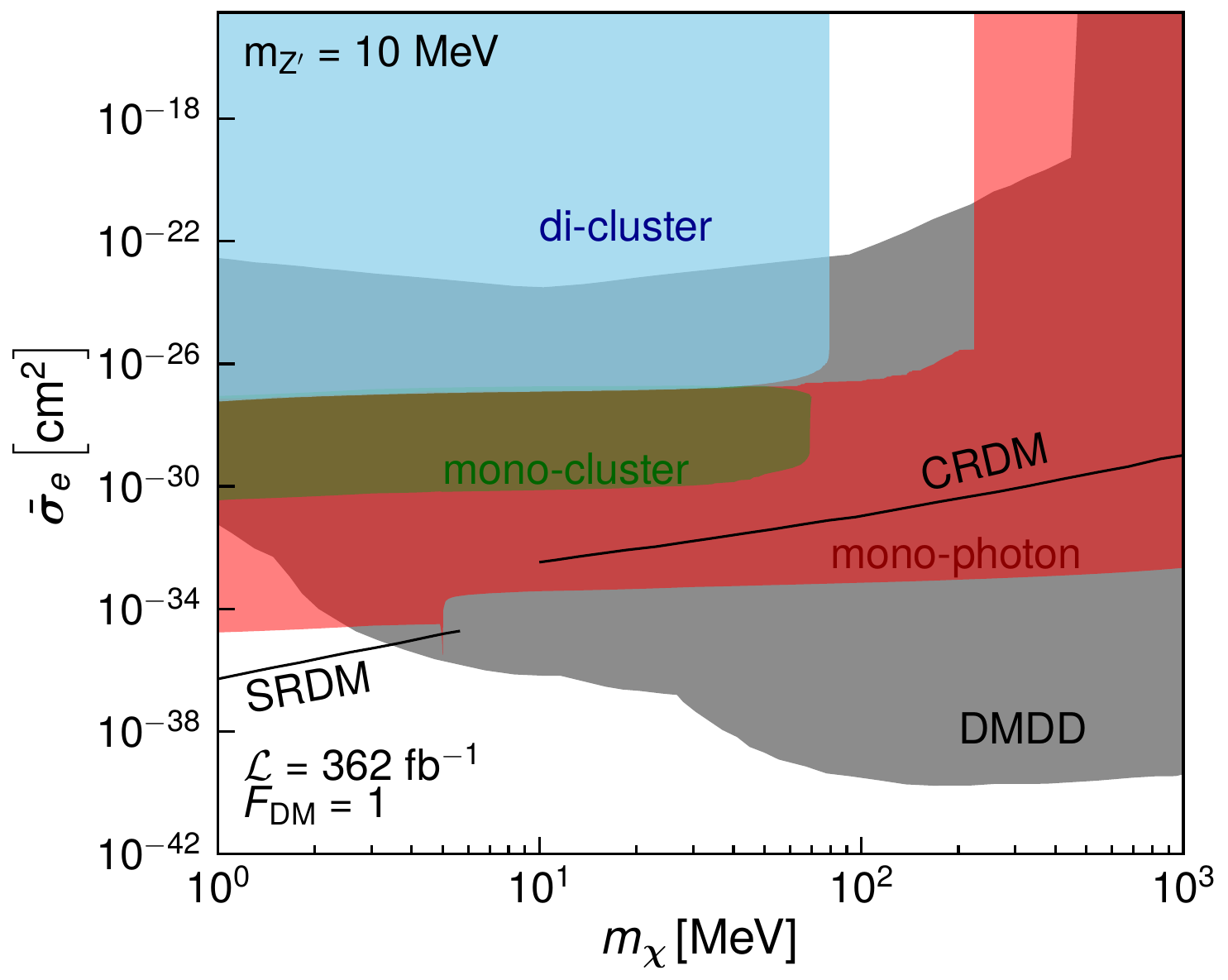}
\includegraphics[width=0.45 \textwidth]{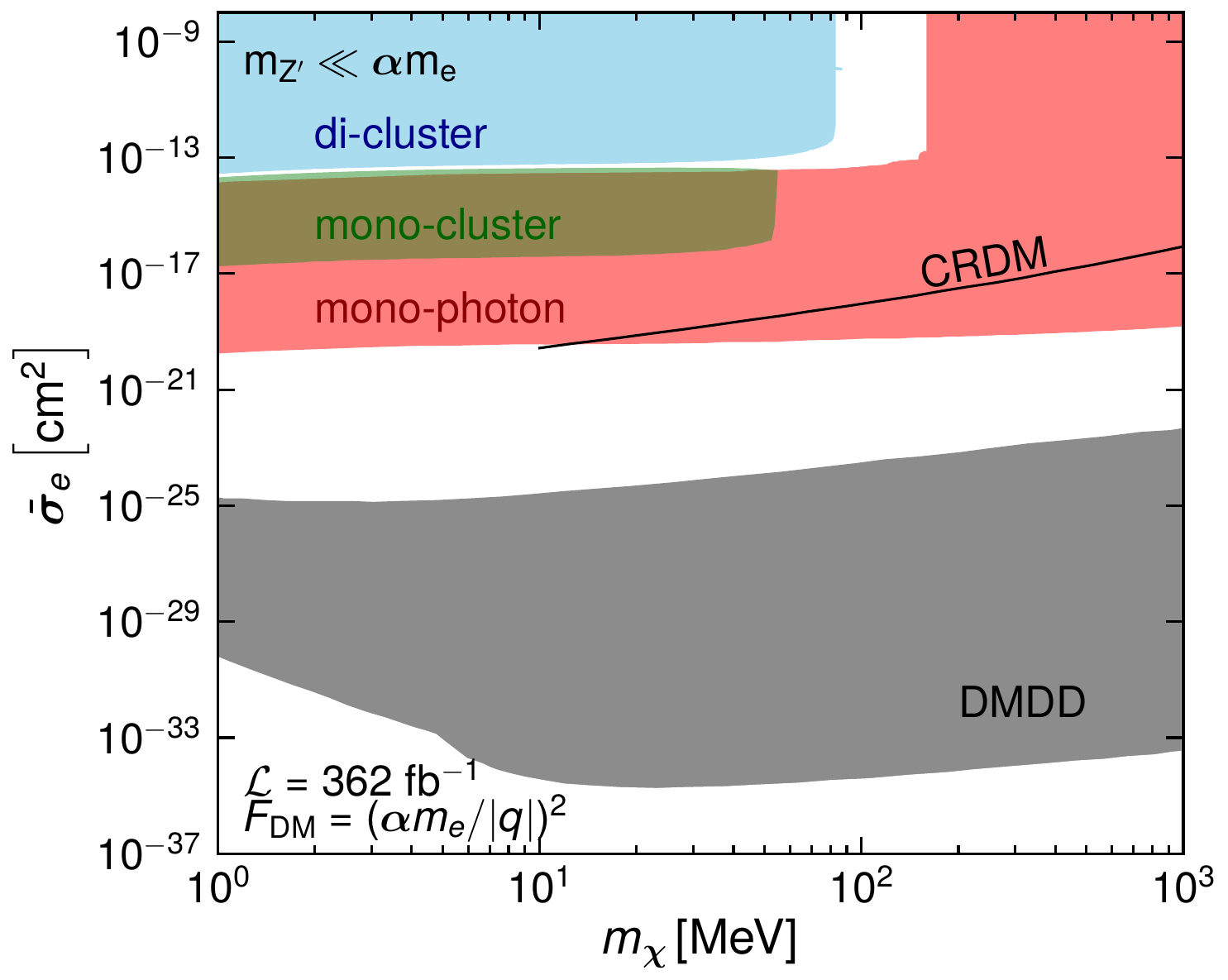}
\caption{
Belle II 90\% C.L.\ sensitivity (with 362 fb$^{-1}$ of data) 
on the DM-electron reference cross section in the vector mediator case: 
$m_{Z'}=10$ MeV (left), 
where the DM form factor is $F_{\rm DM}=1$;  
and $m_{Z'}\ll \alpha m_e$ (right), 
where $F_{\rm DM}=(\alpha m_e/|{\bf q}|)^2$. 
Belle II constraints for three DM channels 
are shown as shaded regions: 
(1) di-cluster (blue), 
(2) mono-photon (red), 
(3) mono-cluster (green). 
The gray shaded regions indicate the DMDD constraints: 
the lower boundary on the left panel is
obtained by combining constraints from 
Xenon10 \cite{Essig:2017kqs},
Xenon1T \cite{Aprile:2019xxb}, 
and SENSEI \cite{SENSEI:2020dpa}; 
the upper boundary (ceiling) on the left panel 
and the exclusion region on the right panel 
are adopted from Ref.~\cite{Emken:2019tni} (with electronic stopping only).
Other constraints are shown as upper limits only (black solid curves): 
Xenon1T limits on boosted DM from solar reflection (denoted as ``SRDM'' here) \cite{An:2017ojc}; 
and Super-K limits on cosmic ray boosted DM (denoted as ``CRDM'' here) 
\cite{Dent:2020syp} 
for mediator masses at 10 MeV (left panel) and eV (right panel).
}
\label{fig:CSeSIMP}
\end{centering}
\end{figure*}

\section{Comparison with DMDD}
\label{sec:refXsec}

In this section we compare the Belle II constraints with DMDD 
constraints (also neutrino experiments) 
that are due to electron recoils.
The Belle II constraints consist of the limits 
from three channels: 
di-cluster, 
mono-cluster, 
and mono-photon. 
We consider the vector mediator model, 
given in Eq.~\eqref{eq:L:V},
as the 
benchmark model in this section. 
To compare constraints with DMDD,
we compute the reference cross section
\cite{Essig:2011nj, Essig:2015cda}
\be
\bar{\sigma}_{e} \simeq  
\left. 
\frac{(g_e^V g_{\chi}^V)^2 \mu_{\chi e}^2}{\pi (m_{Z'}^2 + |{\bf q}|^2)^2 } \right|_{|{\bf q}| = \alpha m_e}, 
\label{eq:ref:xsec}
\ee 
where 
$m_{Z'}$ is the mediator mass, 
and $|{\bf q}|=\alpha m_e$ is the 
magnitude of the mediator's 3-momentum  
evaluated at the typical atomic scale.

Although Belle II constraints on the product of the couplings, 
$g_\chi g_e$, 
do not exhibit a significant dependence on the mediator mass 
across most of the parameter space analyzed in this study, 
the DMDD constraints are very 
sensitive to the mediator mass. 
This is evident in the reference cross section 
defined in Eq.~\eqref{eq:ref:xsec}. 
To properly 
interpret the Belle II constraints 
in the context of DMDD experiments and other experimental constraints that rely on the DM-electron scattering process, 
we consider two distinct scenarios for the mediator mass: 
the ultralight mediator case, namely $m_{Z'} \ll \alpha m_e$, 
and the $m_{Z'}=10$ MeV case.  
%%%
In the $m_{Z'}=10$ MeV case, one has 
\be
\bar{\sigma}_{e} \simeq  \frac{(g_e^V g_{\chi}^V)^2 \mu_{\chi e}^2}{\pi m_{Z'}^4 }. 
\ee 
In the ultralight mediator case, one has  
\be
\bar{\sigma}_{e} \simeq  
\frac{(g_e^V g_{\chi}^V)^2 \mu_{\chi e}^2}{\pi (\alpha m_e)^4  }.
\ee  
The q-dependence of the DM-electron cross 
section can be encoded in the DM form factor 
\cite{Essig:2011nj, Essig:2015cda}. 
These two types of mediators 
have distinct DM from factors, 
which are $F_{\rm DM} (q) = \left(\alpha m_e/|{\bf q}|\right)^2$ 
and $F_{\rm DM} (q) =1$ 
for the 
$m_{Z'} \ll \alpha m_e $ and $m_{Z'}=10 ~\rm MeV$ cases, respectively. 
\footnote{The expressions of $\bar{\sigma}_e$ 
and of the DM form factor 
$F_{\rm DM} (q)$ are given in appendix \ref{sec:CSequation}, 
for the four different mediators.}

We compare the Belle II constraints (using the 
current data of $\mathcal{L}=362 ~ \rm fb^{-1}$) with 
the constraints from DMDD 
and neutrino experiments, 
for the $m_{Z'}=10~\rm MeV$ and $m_{Z'} \ll \alpha m_e$ cases  
in the left and right panels of Fig.~(\ref{fig:CSeSIMP}), 
respectively.

In the $m_{Z'}=10~\rm MeV$ case 
(the left panel figure of Fig.~(\ref{fig:CSeSIMP})), 
the lower boundary of the DMDD exclusion region 
consists of constraints from 
Xenon10 \cite{Essig:2017kqs}, 
Xenon1T \cite{Aprile:2019xxb}, and 
SENSEI \cite{SENSEI:2020dpa}. 
There are also constraints on boosted-DM, including 
Super-K limits on cosmic rays boosted DM \cite{Dent:2020syp}, 
and Xenon1T limits on DM due to solar reflection \cite{An:2017ojc}. 
We find that for the $m_{Z'}=10~\rm MeV$ case, 
the lower boundary of the Belle II mono-photon constraints 
is higher than the combined constraints from 
DMDD/neutrino experiments. 
We note that, 
however, 
the mono-photon constraints can easily surpass the 
DMDD constraints for a mediator with a mass larger than 10 MeV 
\cite{Liang:2021kgw}. 
This is because the Belle II constraints on $\bar{\sigma}_e$ 
are proportional to $1/m_{Z'}^4$ in the mass range of  
$\alpha m_e \ll m_{Z'} \ll \sqrt{s} = 10.58$ GeV.

Ceilings also exist in DMDD constraints. 
This is because DM has to penetrate the overburden 
(including both rock and the atmosphere) 
of the underground DMDD labs. 
If DM has a sufficiently large interaction cross section 
with SM particles, it gets absorbed before reaching 
the underground detectors. 
\footnote{Experiments conducted at the top of the atmosphere or on Earth's surface, 
such as XQC \cite{McCammon:2002gb,Erickcek:2007jv},
CSR \cite{CRESST:2017ues}, 
and RRS \cite{Rich:1987st}, 
provide a more promising avenue for investigating strongly-interacting DM. 
See also Ref.~\cite{Cantatore:2020obc} 
for a proposed experiment in the upper stratosphere aimed at mitigating the overburden. 
Recent analyses of constraints from these experiments on the interaction cross section 
between nuclei and strongly-interacting DM can be found in 
Refs.~\cite{Davis:2017noy,Kavanagh:2017cru,Mahdawi:2017cxz,Hooper:2018bfw,Emken:2018run,Mahdawi:2018euy,Xu:2020qjk,McKeen:2022poo,Li:2022idr,Bramante:2022pmn}.} 
This also applies to constraints from 
neutrino experiments such as Super-K, MiniBooNE, and DUNE 
\cite{Dent:2020syp,Cappiello:2019qsw}.
Ref.~\cite{Emken:2019tni} has analyzed 
the DMDD ceilings for DM-electron interactions, 
and found that the ceilings occur at 
$\bar{\sigma}_e \simeq 10^{-22} (10^{-25}) ~\rm cm^2$ 
in the case where the DM form factor is 
$F_{\rm DM} ({\bf q})=1$ ($F_{\rm DM}({\bf q})= (\alpha m_e /|{\bf q}|)^2$). 
As shown in the left panel figure of Fig.~(\ref{fig:CSeSIMP}), 
for the $m = 10 ~\rm MeV$ case, Belle II 
is capable of ruling out almost 
the entire parameter space above 
the DMDD ceiling, 
except the narrow mass window of 
$\sim$(100-200) MeV.
\footnote{This narrow mass window can become even 
narrower or disappear if multiple scatterings are  
properly taken into account, as discussed in 
section \ref{sec:CPBelle2}.}

In the $m_{Z'} \ll \alpha m_e$ case, 
the Belle II constraints probe a completely 
different parameter space compared to DMDD, 
as shown in the right panel figure of Fig.~(\ref{fig:CSeSIMP}). 
The lower boundary of the mono-photon 
exclusion region is approximately four orders of magnitude 
larger than the ceiling of the DMDD exclusion region. 
Notably, the parameter region of 
$10^{-25} {~\rm cm^2} \lesssim \bar{\sigma}_e \lesssim 10^{-20} {~\rm cm^2}$ for MeV-GeV DM is currently allowed, 
in the ultralight mediator case.
\footnote{We note that the CRDM limit given by 
Ref.~\cite{Dent:2020syp} for the eV mediator mass 
appears above the DMDD ceiling 
analyzed by Ref.~\cite{Emken:2019tni}. 
This might be due to the overlooked attenuation effects 
from the atmosphere and Earth.}

\section{BSM particles from PBH evaporation}
\label{sec:PBH}

\begin{figure}[htbp]
\begin{centering}
\includegraphics[width=0.45 \textwidth]{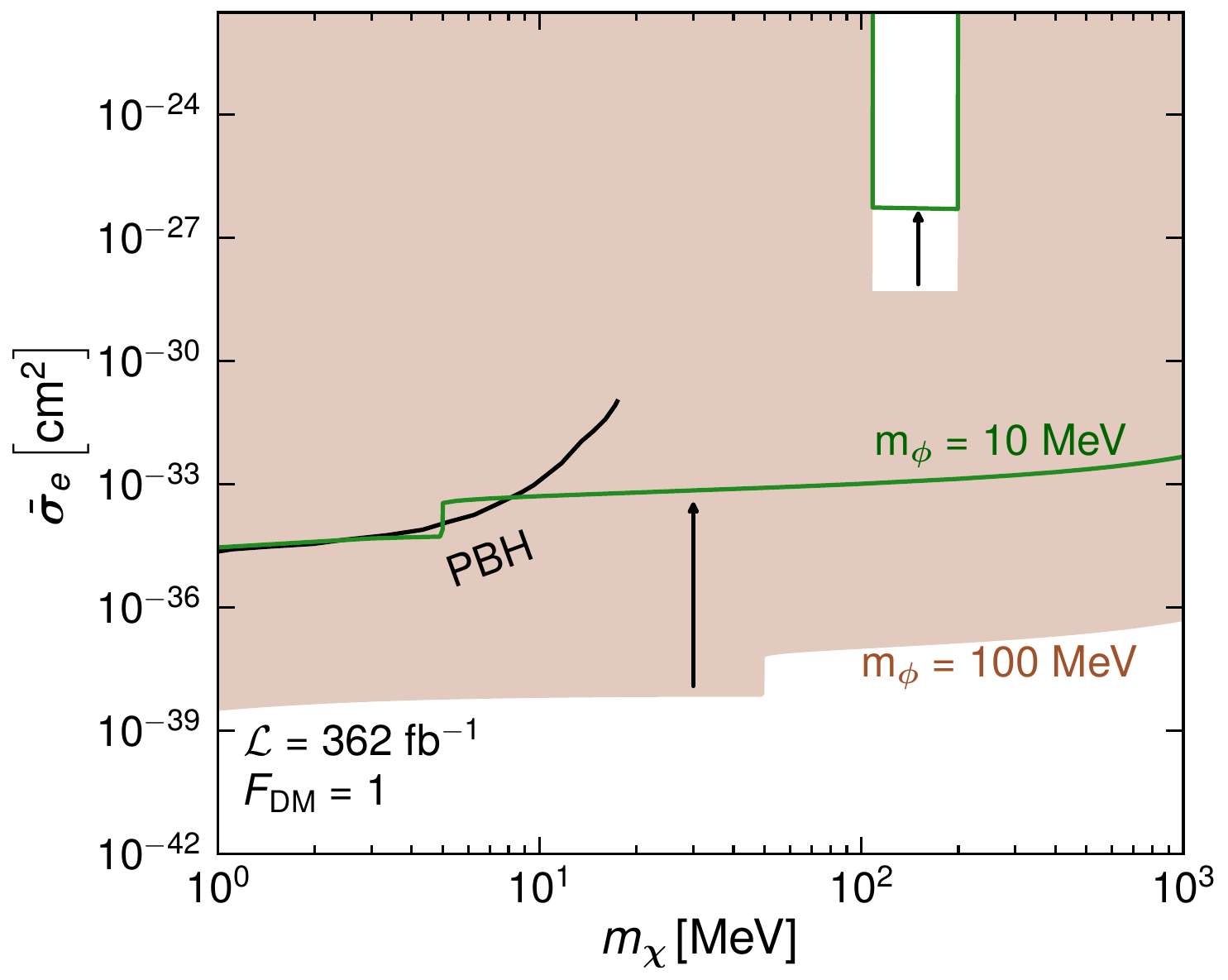}
\caption{Belle II 90\% C.L.\ sensitivity 
(combining constraints from 
mono-photon, mono-cluster, and di-cluster channels)
on the reference cross section 
of the $\chi$-e interaction  
in the scalar mediator case: 
$m_{\phi}=100$ MeV (brown shaded region) 
and $m_{\phi}=10$ MeV (green curve). 
We use 362 fb$^{-1}$ of data for Belle II and 
$F_{\rm DM}=1$ for the DM form factor.   
Super-K limits on $\chi$ from  
PBH evaporation are 
also shown, 
for the scalar EFT operator 
$\bar \chi \chi \bar e e /\Lambda^2$ 
with $M_{\rm PBH}=7.9\times 10^{14}$ g \cite{Calabrese:2022rfa}.}
\label{fig:CSeSIMPS}
\end{centering}
\end{figure}

Light BSM particles with a mass below 
the Hawking temperature can be produced in PBH,
leading to detectable signals at terrestrial experiments, 
if they have particle interactions with 
the SM sector 
\cite{Calabrese:2021src,Li:2022jxo, Calabrese:2022rfa}. 
Ref.~\cite{Calabrese:2022rfa} analyzed the XENON1T and Super-K constraints 
on PBH-evaporated BSM particles that interact with electrons, 
for an EFT interaction of $\bar \chi \chi \bar \ell \ell /\Lambda^2$, 
where $\ell$ denotes the SM leptons. 
Such an EFT operator can be obtained by integrating out 
a scalar mediator with a mass above the typical energy of PBH, 
which is at most $\sim 10$ MeV \cite{Calabrese:2022rfa}. 
To compare Belle II constraints with PBH constraints, 
we thus consider scalar mediator models with mediator masses  
of 10 MeV and 100 MeV, as shown in Fig.~(\ref{fig:CSeSIMPS}).\footnote{We note that 
for the 100 MeV case, the mono-photon ceiling near 100 MeV occurs 
at $g_eg_\chi\sim 4\pi$.}
The Belle II constraints consist of those from 
the mono-photon, mono-cluster, and di-cluster channels. 
We find that the Belle II sensitivity with the current data 
(362 fb$^{-1}$) surpasses the PBH constraint by more than 
four orders of magnitude, for the mediator mass of $m_{\phi}=100$ MeV. 
If the mediator mass is lowered to 10 MeV, 
both the floor and the ceiling of the 
Belle II constraints move upward 
(as indicated by the black arrows in Fig.~(\ref{fig:CSeSIMPS})); 
the resulting Belle II sensitivity is  
at the same level of the PBH constraints. \footnote{If the mediator mass is further lowered to 1 MeV, 
the lower boundary of the Belle II constraints 
on the reference cross section $\bar\sigma_e$ 
will be shifted upward  
by roughly another four orders of magnitude. 
We do not consider this case in Fig.~(\ref{fig:CSeSIMPS}) 
because the PBH constraints 
that are analyzed for an EFT operator 
might be modified significantly 
in the case where the mediator mass is comparable 
or even lower than the Hawking temperature.}

\section{summary}
\label{sec:summary}

This study explores the potential of Belle II experiment to detect 
strongly-interacting DM that couples to electrons.
As the interaction strength between DM and electrons increases, 
DM can no longer evade detection 
but instead deposit significant energy in the calorimeter, 
leading to ECL clusters. 
Thus, for strongly-interacting DM, the mono-photon channel 
becomes invalid. Instead, the visible signatures in the 
ECL, including di-cluster and mono-cluster channels, 
assume significance.

To illustrate the sensitivity of Belle II to strongly-interacting DM, we consider several light mediator models with mass less than 10 MeV, including the vector mediator, 
the axial-vector mediator, the scalar mediator, and the pseudo-scalar mediator.
We analyze constraints on 
such mediators from various experiments, 
including
electron $g-2$, 
electron beam dump experiments,
electron-positron colliders, 
and {M\o ller} scattering.
We find that these experiments impose a strong 
constraint on the coupling between the light mediator 
and electrons. 
Nonetheless, there is significant parameter space  
allowed: $g_e \lesssim 10^{-4}$ for mediator mass $\sim 10$ MeV. 
We note that couplings to electrons can be substantial if there is more than one mediator.

We calculate the exclusion regions for the three different channels 
at Belle II: 
di-cluster, mono-cluster, and mono-photon. 
While the mono-cluster and mono-photon channels exhibit both 
a lower boundary and an upper boundary (ceiling), 
the di-cluster channel is characterized solely by a lower boundary. 
Remarkably, we find that the lower boundaries of the di-cluster channel 
typically reside below the upper boundaries of the mono-photon channel. 
Therefore, the integration of these two channels 
proves effective in exploring the entire parameter space 
above the lower boundaries of the mono-photon exclusion region.

We further compare Belle II constraints 
with constraints from DMDD/neutrino experiments, 
as well as PBH constraints.
We find that 
the di-cluster channel at Belle II
can probe  
the parameter space above the ceilings of DMDD constraints
for the vector mediator case with $m_{Z'}=10 ~\rm MeV$; 
similar conclusions can be extended to mediator models 
with $10 {~\rm MeV} <  m_{Z'} \ll \sqrt{s}$.
For the $m_{Z'}\ll \alpha m_e$ case, however, 
the parameter space probed by Belle II 
is well seperated from that probed by DMDD experiments.
We also find that Belle II constraints on light BSM particles 
are stronger than PBH constraints, 
for the scalar mediator 
in the mass range of $\sim(10-100)$ MeV.

\begin{acknowledgments}

We thank 
Yonglin Li, 
Wenxi Lu, 
and Zicheng Ye 
for correspondence and discussions. 
The work is supported in part by the 
National Natural Science Foundation of China under Grant 
Nos.\ 12275128, 12147103, and 12347121, 
and by the
Guangdong Major Project of Basic and Applied Basic Research 
No.\ 2020B0301030008.

\end{acknowledgments}

\appendix

\begin{widetext}

\section{Four models and their cross sections}
\label{sec:CSequation}

In this section we collect cross section 
formulas for different models 
with the following four types of mediators: 
(1) vector, 
(2) axial-vector, 
(3) scalar, 
and 
(4) pseudo-scalar. 
The Lagrangian of these four models are given in 
Eqs.~(\ref{eq:L:V}-\ref{eq:L:P}). 
For each model, 
we provide the differential cross sections 
for the following three processes: 
(1) $e^+ e^- \to \chi \bar{\chi}$,
(2) $e^+ e^- \to \chi \bar{\chi} \gamma$,
and (3) $\chi e^- \to \chi e^-$.

\subsection{DM production cross section}

The differential cross sections 
of the $e^+e^-\to\chi\bar{\chi}$ process for the four mediators are given by
\begin{align}
\frac{d\sigma_{\chi\bar{\chi}}}{dz_\chi^* }  
&=\frac{ (g_e^V g_{\chi}^V)^2\beta_\chi s \left [ 
 z_\chi ^{*2}  \beta_e^2 \beta_\chi^2+3-\beta_e^2-\beta_\chi^2 
\right]}{32\pi\beta_e\left[\left(s-m_{Z^{\prime}}^{2}\right)^{2}+m_{Z^{\prime}}^{2}
\Gamma_{Z^{\prime}}^{2}\right] },\\
\frac{d\sigma_{\chi\bar{\chi}}}{dz_\chi^* }  
&=\frac{ (g_e^A g_{\chi}^A)^2 \beta_\chi s \left [ \beta_e^2\beta_\chi^2(z_\chi ^{*2} +1)+(1-\beta_e^2)(1-\beta_\chi^2)  
\right]}{32\pi\beta_e  \left[\left(s-m_{Z^{\prime}}^{2}\right)^{2}+m_{Z^{\prime}}^{2}
\Gamma_{Z^{\prime}}^{2}\right] },\\
\frac{d\sigma_{\chi\bar{\chi}}}{dz_\chi^* }  
&=\frac{  (g_e^S g_{\chi}^S)^2\beta_e \beta_\chi^3 s }{32\pi\left[\left(s-m_{\phi}^{2}\right)^{2}+m_{\phi}^{2}
\Gamma_{\phi}^{2}\right] },\\
\frac{d\sigma_{\chi\bar{\chi}}}{dz_\chi^* }  
&=\frac{  (g_e^P g_{\chi}^P)^2 \beta_\chi s }{32\pi\beta_e 
\left[ \left( s-m_{\phi}^{2}\right)^{2}+m_{\phi}^{2}
\Gamma_{\phi}^{2} \right] },
\label{eq:ann}
\end{align}
where 
$z_\chi^* = \cos \theta_\chi^*$ with $
\theta_\chi^*$ being the polar angle of $\chi$ in the c.m.\ frame, 
$s$ is the square of the center-of-mass energy, 
$m_{Z^\prime}$ ($\Gamma_{Z^\prime}$) is the mass 
(decay width) of the spin-1 mediator, 
$m_{\phi}$ ($\Gamma_{\phi}$) is the mass 
(decay width) of the spin-0 mediator, 
$\beta_{\chi}=\sqrt{1-4m_{\chi}^2/s}$ 
with $m_\chi$ being the DM mass, 
and 
$\beta_{e}=\sqrt{1-4m_{e}^2/s}$ 
with $m_e$ being the electron mass. 
Note that one can take $\beta_e = 1$ 
at Belle II.

\subsection{Mono-photon cross section}

The differential cross sections of the mono-photon signal 
from the $e^+e^-\to\chi\bar{\chi} \gamma$ process 
(with the phase space of DM particles  
integrated out)
are given by,
\bea
\frac{d \sigma_{\chi\bar{\chi}\gamma }}{d E_{\gamma}^*  d z_{\gamma}^* }  &=& 
\frac{\alpha (g_e^V g_{\chi}^V)^2
s_{\gamma}^{2} 
}{6 \pi^{2} s E_{\gamma}^* 
\left[\left(s_{\gamma}-m_{Z^{\prime}}^{2}\right)^{2}+m_{Z^{\prime}}^{2}
\Gamma_{Z^{\prime}}^{2}\right] }  
\sqrt{1-4\frac{m_{\chi}^2}{s_\gamma}}
\left[\frac{1}{{1-z_{\gamma}^{*2}}}+\frac{E_\gamma^{*2}}{s_\gamma}
\frac{1+z_{\gamma}^{*2}}{1-z_\gamma^{*2} }\right]
\left(1+2 \frac{m_{\chi}^2}{s_\gamma}\right),\\
\frac{d \sigma_{\chi\bar{\chi}\gamma}}{d E_{\gamma}^*  d z_{\gamma}^* }  &=& 
\frac{\alpha (g_e^A g_{\chi}^A)^2
s_{\gamma}^{2} 
}{6 \pi^{2} s E_{\gamma}^*
\left[\left(s_{\gamma}-m_{Z^{\prime}}^{2}\right)^{2}+m_{Z^{\prime}}^{2}
\Gamma_{Z^{\prime}}^{2}\right] }  
\sqrt{1-4\frac{m_{\chi}^2}{s_\gamma}}
\left[\frac{1}{{1-z_{\gamma}^{*2}}}+\frac{E _\gamma^{*2}}{s_\gamma}
\frac{1+z_{\gamma}^{*2}}{1-z_\gamma^{*2} }\right]
\left(1-4 \frac{m_{\chi}^2}{s_\gamma}\right),\\
\frac{d \sigma_{\chi\bar{\chi}\gamma}}{d E_{\gamma}^*  d z_{\gamma}^* }  &=&
\frac{\alpha (g_e^S g_{\chi}^S)^2
 \left(s_\gamma^2+s^2\right) s_{\gamma}
}{16 \pi^{2} s^{2} E_{\gamma}^* \left(1-z_\gamma^{*2} \right)
\left[\left(s_{\gamma}-m_{\phi}^{2}\right)^{2}+m_{\phi}^{2}
\Gamma_{\phi}^{2}\right] } 
\sqrt{1-4\frac{m_{\chi}^2}{s_\gamma}}
\left(1-4 \frac{m_{\chi}^2}{s_\gamma}\right),
\\
\frac{d \sigma_{\chi\bar{\chi}\gamma}}{d E_{\gamma}^*  d z_{\gamma}^* }  &=& 
\frac{\alpha (g_e^P g_{\chi}^P)^2
 \left(s_\gamma^2+s^2\right)s_{\gamma}
}{16 \pi^{2} s^{2} E_{\gamma}^* \left(1-z_\gamma^{*2} \right)
\left[\left(s_{\gamma}-m_{\phi}^{2}\right)^{2}+m_{\phi}^{2}
\Gamma_{\phi}^{2}\right] } 
\sqrt{1-4\frac{m_{\chi}^2}{s_\gamma}},
\label{eq:monoZp}
\eea
where 
$E_\gamma^* $ is the energy of the photon in the c.m.\ frame,
$z_\gamma^* =\cos\theta_\gamma^* $ with $\theta_\gamma^* $ being the angle of the photon 
in the c.m.\ frame, 
$s_\gamma=s-2\sqrt{s}E_\gamma^* $.
Here, we have used $\beta_e=1$.

\subsection{DM-electron scattering cross section}
\label{app:sca}

The differential scattering cross sections of 
the $ \chi e^- \to  \chi e^-$ process for four mediator models are given by
\bea
\label{eq:sca}    
\frac{d\sigma_{\chi e}}{dt_{\chi e}}&&=\frac{(g_e^V g_{\chi}^V)^2\left(8m_e^2E_\chi^2+2s_{\chi e}t_{\chi e}+t^2_{\chi e}\right)}{32\pi m_e^2\left((m_{Z'}^2-t_{\chi e})^2+m_{Z'}^2\Gamma_{Z'}^2\right)(E_\chi^2-m_\chi^2)},\\
\frac{d\sigma_{\chi e}}{dt_{\chi e}}&&=\frac{(g_e^A g_{\chi}^A)^2\left(8m_e^2(E_\chi^2+2m_\chi^2)-2t_{\chi e}(s_{\chi e}-4m_e E_\chi)+t^2_{\chi e}\right)}{32\pi m_e^2\left((m_{Z'}^2-t_{\chi e})^2+m_{Z'}^2\Gamma_{Z'}^2\right)(E_\chi^2-m_\chi^2)},\\
\frac{d\sigma_{\chi e}}{dt_{\chi e}}&&=\frac{(g_e^S g_{\chi}^S)^2\left(4m_e^2-t_{\chi e}\right)\left(4m_\chi^2-t_{\chi e}\right)}{64\pi m_e^2\left((m_{\phi}^2-t_{\chi e})^2+m_{\phi}^2\Gamma_{\phi}^2\right)(E_\chi^2-m_\chi^2)},\\
\frac{d\sigma_{\chi e}}{dt_{\chi e}}&&=\frac{(g_e^P g_{\chi}^P)^2t_{\chi e}^2}{64\pi m_e^2\left((m_{\phi}^2-t_{\chi e})^2+m_{\phi}^2\Gamma_{\phi}^2\right)(E_\chi^2-m_\chi^2)},
\eea
where $t_{\chi e}$ and $s_{\chi e}$ are the Mandelstam variables of 
the DM-electron scattering process,
which are given by
\bea
s_{\chi e}&=&2 m_e E_\chi+m_e^2+m_\chi^2,\\
t_{\chi e}&=&-2m_eE_r,
\eea
where $E_\chi$ and $E_r$
are the energy of the initial state DM 
and the recoil energy of the electron, respectively,
in the lab frame (the rest frame of the initial state electron).
If one is interested in electron recoil events above  
the energy threshold of $E_r^{\rm th}$, 
the total cross section can be obtained by
\be
\sigma_{\chi e}^v(E_\chi,E_r^{\rm th})=\int_{{ t^{\rm min}_{\chi e}}}^{-2 m_e E_r^{\rm  th}} \frac{d\sigma_{\chi e}}{d t_{\chi e}} d t_{\chi e},
\label{eq:crossex}
\ee
where 
\be
t^{\rm min}_{\chi e} = - \frac{1}{s_{\chi e}} 
\left[s_{\chi e}-(m_\chi+m_e)^2\right] \left[ s_{\chi e}-(m_\chi-m_e)^2 \right].
\label{eq:txemin}
\ee

\subsection{Reference cross section in DMDD}

The reference cross section for the DM-electron interaction in DMDD experiments
is defined by \cite{Essig:2011nj, Essig:2015cda}
\be 
\bar{\sigma}_{e} \equiv \frac{\mu_{\chi e}^{2}}{16 \pi m_{\chi}^{2} m_{e}^{2}} 
 \left. \overline{\left|\mathcal{M}_{\chi e}({\bf q})\right|^{2}}
 \right|_{|{\bf q}| = \alpha m_{e}}, 
 \label{eq:trans}
\ee 
where $\mu_{\chi e}=m_\chi m_e/(m_\chi+m_e)$ is the reduced mass.
Here, we have assumed that
both DM and electron are 
non-relativistic,  
and the $q$-dependence is only in the matrix element ${\cal M}_{\chi e}$, 
which can be factorized as
$\overline{\left|\mathcal{M}_{\chi e}({\bf q})\right|^{2}}=
\left. \overline{\left|\mathcal{M}_{\chi e}({\bf q})\right|^{2}} \right|_{|{\bf q}| = \alpha m_{e}}
\left|F_{\mathrm{DM}}({\bf q})\right|^{2}$.
We provide the reference cross sections for the four mediator models as follows
\begin{equation}
  \bar{\sigma}_{e} \simeq \left.  C \frac{(g_e g_{\chi})^2 \mu_{\chi e}^2}{\pi (m^2+|{\bf q}|^2)^2}\right|_{|{\bf q}| = \alpha m_e},   
  \label{eq:refxsec}
\end{equation}
where $C=\{1,3,1,|{\bf q}|^4/(16 m_e^2m_\chi^2)\}$
for the vector mediator, the axial-vector mediator, 
the scalar mediator, and the pseudo-scalar mediator, respectively. 
For the vector, axial vector, and scalar cases,
the form factor is found to be $F_{\rm DM } =1 $ ($F_{\rm DM}=\left(\alpha m_e/|{\bf q}|\right)^2$)
when $m \gg \alpha m_e$ ($m \ll \alpha m_e$).
For the pseudo-scalar case,
one has 
$F_{\rm DM}=|{\bf q}|^2/(\alpha m_e)^2$ ($F_{\rm DM}=|{\bf q}|/(\alpha m_e)$)
when $m \gg \alpha m_e$ ($m \ll \alpha m_e$).

\subsection{Decay width}

At the tree level, 
the mediator ($Z'$ or $\phi$) 
can decay into either $ \chi \bar \chi$ or $e^+ e^-$, 
when kinematically allowed.
Thus, we compute the total decay width via
\begin{equation}
\Gamma_{\rm total}= 
\Gamma_{ \chi \bar\chi} 
+ \Gamma_{e^+ e^-}, 
\label{eq: Zbosondecaywidth}
\end{equation}
where the subscript denotes the final state. 
The invisible decay widths for the four mediator models are given by
\begin{align}
\Gamma_{\chi \bar\chi} = 
\Gamma\left(Z^{\prime} \rightarrow \chi \bar{\chi}\right)&=
\frac{\left(g_{\chi}^{V}\right)^{2}m_{Z^{\prime}}}{12 \pi} \sqrt{1-4 \frac{m_{\chi}^{2}}{m_{Z^{\prime}}^{2}}}
\left(1+2 \frac{m_{\chi}^{2}}{m_{Z^{\prime}}^{2}}\right)
,\\
\Gamma_{ \chi \bar \chi} = 
 \Gamma\left(Z^{\prime} \rightarrow \chi \bar{\chi}\right)&=
\frac{\left(g_{\chi}^A\right)^{2}m_{Z^{\prime}}}{12 \pi} \sqrt{1-4 \frac{m_{\chi}^{2}}{m_{Z^{\prime}}^{2}}}
\left(1-4 \frac{m_{\chi}^{2}}{m_{Z^{\prime}}^{2}}\right),
\\
\Gamma_{\chi \bar\chi} = 
 \Gamma\left(\phi \rightarrow \chi \bar{\chi}\right)&=
\frac{\left(g_{\chi}^{S}\right)^{2}m_{\phi}}{8 \pi} \sqrt{1-4 \frac{m_{\chi}^{2}}{m_{\phi}^{2}}}
\left(1-4 \frac{m_{\chi}^{2}}{m_{\phi}^{2}}\right),\\
\Gamma_{\chi \bar \chi} = 
 \Gamma\left(\phi \rightarrow \chi \bar{\chi}\right)&=
\frac{\left(g_{\chi}^{P}\right)^{2}m_{\phi}}{8 \pi} \sqrt{1-4 \frac{m_{\chi}^{2}}{m_{\phi}^{2}}}.
\label{eq: ZtoDMwidth}  
\end{align}
The visible decay width 
$\Gamma_{e^+ e^-}$ can be obtained 
from $\Gamma_{ \chi \bar \chi}$, 
by substituting the couplings and mass of DM 
with those of electrons.

\end{widetext}

\normalem
\bibliography{ref.bib}
\bibliographystyle{utphys28mod}

\end{document}